\documentclass[12pt,a4paper]{JHEP3}
\usepackage{graphicx}
\usepackage{subfigure}
\newcommand{\beq}{\begin{equation}}
\newcommand{\eeq}{\end{equation}}
\newcommand{\baq}{\begin{eqnarray}}
\newcommand{\eaq}{\end{eqnarray}}

\newcommand{\mc}{\mathcal{}}

\title{The Subdominant Curvaton}
\author{Kari Enqvist$^{a,b,}\footnote{E-mail: kari.enqvist@helsinki.fi}$,
Sami Nurmi$^{c,}\footnote{E-mail:
s.nurmi@thphys.uni-heidelberg.de}$, Gerasimos
Rigopoulos$^{b,}$\footnote{E-mail:
gerasimos.rigopoulos@helsinki.fi}, Olli
Taanila$^{a,b,}$\footnote{E-mail: olli.taanila@iki.fi},
 Tomo Takahashi$^{d,}$\footnote{E-mail: tomot@cc.saga-u.ac.jp}
\\
$^a$Physics Department, FIN-00014 University of Helsinki, Finland;\\
$^b$Helsinki Institute of Physics, FIN-00014 University of Helsinki,
Finland;\\
$^c$Institute for Theoretical Physics, University of Heidelberg,
69120 Heidelberg, Germany;\\
$^d$Department of Physics, Saga University, Saga 840-8502, Japan }

\abstract {We present a systematic study of the amplitude of the
primordial perturbation in curvaton models with self-interactions,
treating both renormalizable and non-renormalizable interactions. In
particular, we consider the possibility that the curvaton energy
density is subdominant at the time of the curvaton decay. We find
that large regions in the parameter space give rise to the observed
amplitude of primordial perturbation even for non-renormalizable
curvaton potentials, for which the curvaton energy density dilutes
fast. At the time of its decay, the curvaton energy density may
typically be subdominant by a relative factor of $10^{-3}$ and still
produce the observed perturbation. Field dynamics turns out to be
highly non-trivial, and for non-renormalizable potentials and
certain regions of the parameter space we observe a non-monotonous
relation between the final curvature perturbation and the initial
curvaton value. In those cases, the time evolution of the primordial
perturbation also displays an oscillatory behaviour before the
curvaton decay.}

 \preprint{HIP}
\begin{document}

\section{Introduction}

In the simplest curvaton mechanism \cite{curvaton}, primordial
perturbations originate from quantum fluctuations of a light scalar
field which gives a negligible contribution to the total energy
density during inflation. This field is called the curvaton
$\sigma$. Inflation is driven by another scalar, the inflaton
$\phi$, whose potential energy dominates the universe. After the end
of inflation, the inflaton decays into radiation. If the
inflationary scale is low enough, $H_{*}\ll 10^{-5}
\sqrt{\epsilon_{*}}$, the density fluctuations of the radiation
component are much below the observed amplitude $\delta\rho/\rho\sim
10^{-5}$ and the fluid is for practical purposes homogeneous. While
the dominant radiation energy scales away as $\rho_{r}\sim a^{-4}$,
the curvaton energy gets diluted at a slower rate, at least for
quadratic curvaton potentials. The curvaton contribution to the
total energy density therefore increases and the initially
negligible curvaton perturbations get imprinted into metric
fluctuations. The standard adiabatic hot big bang era is recovered
when the curvaton eventually decays and thermalizes with the
existing radiation. The mechanism can be seen as a conversion of
initial isocurvature perturbations into adiabatic ones and,
depending on the parameters of the model, is capable of generating
all of the observed primordial perturbation.

The scenario sketched above represents the simplest possible
realization of the curvaton mechanism and a wide range of different
variations of the idea have been studied in the literature. For
example, the inflaton perturbations need not be negligible
\cite{mixed}, there could be several curvatons
\cite{many_curvatons}, the curvaton decay can result into residual
isocurvature perturbations \cite{LUW,isocurvature} and inflation
could be driven by some other mechanism than slowly rolling scalars
\cite{alt_inflation}. Recently it has been pointed out that the
curvaton can decay via a parametric resonance \cite{curvatondecres}
(see also \cite{BasteroGil:2003tj}) with potentially interesting
observational consequences. It is also well known that the
predictions of the curvaton model are quite sensitive to the form of
the curvaton potential. In particular, even small deviations from
the extensively studied quadratic potential can have a significant
effect, at least when considering non-Gaussian effects \cite{kesn,
kett}. However, thus far the predictions of a curvaton scenario with
self-interactions have not been properly studied. This is the aim of
the present paper.

We explore in detail the curvaton scenario with the potential given by
\begin{equation}
V = \frac{1}{2}m^2\sigma^2 + \lambda{\sigma^{n+4}\over M^n} \ ,
\label{curvatonpot}
\end{equation}
where $n$ is an even integer to keep the potential bounded from
below and the interaction term is suppressed by a cut-off scale $M$.
For non-renormalizable operators $n > 0$ we set the cut-off scale to
be the Planck scale, i.e.\ set $M = M_{\rm P}$, and the coupling to
unity, $\lambda = 1$. For the  renormalizable quartic case $n = 0$,
we treat the coupling $\lambda$ as a free parameter. For the rest of
the paper we set Planck mass to unity $M_{\rm P}\equiv(8\pi
G)^{-1/2}= 1$.

Our approach is purely phenomenological and we do not explicitly
connect our discussion to any particle physics model in this work.
Even without a concrete model in mind, the study of the potential
(\ref{curvatonpot}) is however well motivated by generic theoretical
arguments. Indeed, the curvaton should have interactions of some
kind as it eventually must decay and produce Standard Model fields.
The curvaton needs to be weakly interacting to keep the field light
during inflation. This however only implies that the effective
curvaton potential should be sufficiently flat in the vicinity of
the field expectation value during inflation but does not a priori
require the interaction terms in (\ref{curvatonpot}) to be
negligible. Moreover, as the inflationary energy scale is relatively
high, the field can be displaced far from the origin and therefore
feel the presence of higher order terms in the potential. The
interactions could arise either as pure curvaton self-interactions
involving the field $\sigma$ alone or more generically as effective
terms due to curvaton couplings to other (heavy) degrees of freedom
that have been integrated out. An example of a possible physical
setup which could lead to (\ref{curvatonpot}) is given by flat
directions of supersymmetric models that have been suggested as
curvaton candidates \cite{curvaton_flat}. This would lead to a
potential of the form (\ref{curvatonpot}) with typically a
relatively large power for the non-renormalizable operator.

We consider for simplicity a perturbative curvaton decay
characterized by some decay width $\Gamma$ and do not address the
possibility of a non-perturbative decay \cite{curvatondecres}. Our
analysis is numerical and covers both the regions of small and large
interactions in (\ref{curvatonpot}) as compared to the quadratic
part. For small interactions, the deviations from quadratic results
appear mostly when considering beyond leading order properties like
non-gaussianties \cite{kesn, kett}. The opposite case with large
interactions is much less studied, although some work has been done
\cite{dynamics,Huang}, and is therefore of primary interest for the
current work. When the interaction term dominates in
(\ref{curvatonpot}), the curvaton oscillations start in a
non-quadratic potential and the curvaton energy density always
decreases faster than for a quadratic case. For non-renormalizable
interactions, the decrease is even faster than the red-shifting of
the background radiation and the curvaton contribution to the total
energy density is decreasing at the beginning of oscillations.
Consequently, the amplification of the curvaton component is less
efficient than for a quadratic model. For the same values for $m$
and $\Gamma$, the curvaton typically ends up being more subdominant
at the time of its decay than in the quadratic case. This we call
the subdominant curvaton scenario.

Despite the subdominance, the curvaton scenario can yield the
correct amplitude of primordial perturbations as the relative
curvaton perturbations $\delta\sigma_{*}/\sigma_{*}$ produced during
inflation can be much larger than $10^{-5}$. For a quadratic model,
it is well known that the curvaton should make up at least few per
cents of the total energy density at the time of its decay,
$r={\rho_{\sigma}}/{\rho_{r}}\gtrsim\mc{O}(10^{-2})$, in order not
to generate too large non-Gaussianities \cite{LUW,curvaton_ng}. This
bound does not directly apply to the non-quadratic model
(\ref{curvatonpot}) since the dynamics is much more complicated.
Although the subdominant curvaton scenario implies relatively large
perturbations $\delta\sigma_{*}/\sigma_{*}$, the higher order terms
in the perturbative expansion of curvature perturbation can be
accidentally suppressed \cite{kesn,kett}. Although the study of
non-Gaussianities is of great importance, stringent constraints on
model building already come from the amplitude of CMB fluctuations.
Therefore, in the present paper we focus on exploring the parameter
space that produces the observed amplitude for the model
(\ref{curvatonpot}) and leave the study of non-Gaussianities for a
separate publication. As it will turn out, already this first step
does place quite non-trivial constraints on the viable parameter
space.

We find that the observed $10^{-5}$ amplitude of primordial
perturbations can be produced for all the interaction terms,
$n=0,2,4,6$, in (\ref{curvatonpot}) which we consider. To illustrate
the resulting constraints for the model parameters, we represent the
corresponding value of $r=\rho_{\sigma}/\rho_{r}$ at the decay time
as a function of the inflationary scale $H_{*}$ and the curvaton
energy during inflation $r_{*}$. This describes how subdominant the
curvaton ends up being at the time of its decay. For the special
case of quartic self-interactions, $n=0$, we also present a new
analytical approach which can be used to estimate the curvature
perturbation and compare this with our numerical analysis. Our
results imply that, as far as the amplitude of perturbations is
concerned, the interacting model (\ref{curvatonpot}) is compatible
with observations even for large interactions. This completes the
first step in determining the observational status of the model
(\ref{curvatonpot}) and makes the study of the non-Gaussianities
both well motivated and a necessary task to complete the work.

The paper is organized as follows. In Sect.\ 2 we describe the subdominant curvaton model,
the associated constraints, and discuss the qualitative features of the evolution of
the curvaton energy density and the primordial perturbation. Sect.\ 3 contains our
numerical results. In Sect.\ 4 we present an analytical, approximate solution for
the curvature perturbation in the case of a curvaton models which is a combination
of a quartic and quadratic parts. Finally, in Sect.\ 5 we give a discussion of the results.

\section{The subdominant curvaton model}

\subsection{Model constraints}
\label{modelconstr}

We consider a phenomenological curvaton potential
(\ref{curvatonpot}) where $m$ is the curvaton mass and the last term
represents the leading interaction term. For $n=0$ we obtain
marginally renormalizable four-point interaction. Larger exponents
correspond to non-renormalizable operators and for these we set
$\lambda=1$ and assume the associated cut-off scale is given by the
Planck mass $M=M_{\rm P}=1$. We choose to consider even integer
values for $n$ to keep the potential bounded from below. The
exponent values considered in detail in this paper are $n=0,2,4,6$.
For $n>6$ there are no field oscillations in the non-quadratic part
of the potential; as a consequence, are of no great interest.

For the purposes of the present paper we assume for definiteness
that the inflaton decays instantly to radiation so that
$\rho(\phi_{end})=3H_*^2=\rho_r$. After inflation ends, the curvaton
field must also eventually decay. We consider a perturbative
curvaton decay characterized by a constant decay rate $\Gamma$,
although the curvaton could also decay non-perturbatively by way of
a parametric resonance \cite{curvatondecres}. The potential
(\ref{curvatonpot}) then gives rise to a coupled set of equations of
motion in a radiation dominated background:
\begin{eqnarray}
\label{eom1}
 &\ddot{\sigma}ÃÂ + \left(3H+\Gamma\right)\dot{\sigma} + m^2\sigma + \lambda\left(n+4\right) \sigma^{n+3} = 0 &  \nonumber\\
&\dot{\rho_{\rm{r}}} = -4H \rho_{\rm{r}} + \Gamma \dot{\sigma}^2&  \nonumber\\
&3H^2 = \rho_{\rm{r}} + \rho_\sigma  \ , &
\end{eqnarray}
where $\rho_{\rm{r}}$ is the radiation density and  $\rho_\sigma$ the curvaton density.
We denote the relative curvaton energy density by
\begin{equation}
r\equiv{\rho_\sigma\over \rho_r}\ .
\label{defr}
\end{equation}
The time evolution of $r$ and therefore the rate of the
amplification of the curvaton component depends on the order $n$ of
the interaction term in (\ref{curvatonpot}). In addition to $n$, the
curvaton mass $m$, and the decay width $\Gamma$, the solutions to
the equations of motion depend on two initial conditions: the value
of the inflationary scale $H_*$, and the initial curvaton value
$\sigma_{*}$ which can be equivalently expressed in terms of the
initial curvaton energy density $r_*\sim\rho_{\sigma_*}/H_{*}^2$. In
principle, these are all free parameters albeit subject to the
constraint coming from requirement of curvaton subdominance during
inflation $r_*\ll 1$ and masslessness $V''\ll H_{*}^2$. Moreover, we
assume that the curvature perturbations generated during inflation
are negligible and that the evolution of $H_{*}$ is such that it
gives rise to the correct spectral index for the curvaton
perturbations. For example, a single field slow roll inflation with
$H_{*}\ll 10^{-5} \sqrt{\epsilon_*}$ and $\epsilon_{*}\sim 10^{-2}$
satisfies these conditions.

An additional constraint comes from the time of the curvaton decay
which should take place early enough so as not to spoil the success
of the adiabatic hot big bang model. In particular, we require that
the curvaton must produce only curvature and not (large)
isocurvature perturbations. This corresponds to the situation where
radiation and dark matter have both the same perturbation. Hence, we
need to require that the curvaton decays before dark matter
decouples from the primordial plasma. In other words, if dark matter
decouples at the temperature $T_{\rm DM}$, then $\Gamma\gtrsim
H_{\rm DM}$. For illustrative purposes, let us write $\Gamma = g^2m$
with $g\ll 1$, where $g$ is some coupling constant, and assume
$T_{\rm DM}\sim {\cal O}(10)$ GeV, as would be appropriate for LSP
dark matter. Then we should require
\begin{equation}
\Gamma\gtrsim {\cal O}(10)T_{\rm{DM}}^2\sim 10^{-33}\sim
10^{-15}~{\rm GeV},
\label{decaybeforeDM}
\end{equation}
or $g\gtrsim 10^{-9}({\rm TeV}/m)^{1/2}$.  In what follows, we will
consider curvaton masses within the range $m=10^4 \dots 10^{12}$
GeV, so that for us the limit on the coupling constant $g$ varies
between $g\gtrsim 10^{-14}\dots 10^{-10}$, which from a particle
physics point of view does not seem to be excessively restrictive.

Although the initial value for the curvaton, and hence $r_{*}$, is a free parameter constrained in our case by
\[ r_{*}\ll 1 \]
one can obtain an estimate of its likely value based on a stochastic
argument \cite{equilibrium}. Consider the curvaton rolling down its
potential during inflation. The comoving horizon shrinks and all the
curvaton modes for which $k\gtrsim aH$ appear as stochastic noise in
the dynamics, thus creating stochastic perturbations on scales
$k<aH$ as time passes. Note that, unlike for the inflaton in eternal
inflation, for the curvaton there are no quantum "kicks" that would
keep it indefinitely in the quantum regime; this is so because
during inflation the background is fixed by the inflaton to be
approximately deSitter and is almost independent of the subdominant
curvaton. The one-point probability distribution
$P[\sigma(\vec{x}),t]$ for the curvaton obeys a Fokker-Planck
equation
\beq\label{FP}
{\partial P\over\partial t}  = {1\over 3H} {\partial\over\partial \sigma}\left(V_\sigma P\right)
+{H^3\over 8\pi^2}{\partial^2P\over\partial \sigma^2}
\eeq
The first term on the rhs arises from the classical dynamics while the second represents the quantum noise.
Here, by $\sigma(\vec{x})$ we mean the curvaton smoothed on scales larger than the horizon and $\vec{x}$
denotes the position of one such superhorizon patch.

At some point the comoving horizon will equal our observable
universe and after that all the quantum noise of the curvaton will
correspond to the perturbations we observe within our horizon.
Therefore, the classical background value for the curvaton in our
universe will be drawn by the solution to (\ref{FP}) at the time of
horizon crossing for our observable universe. Of course this will
depend on the initial condition for the curvaton at the beginning of
inflation, but, assuming that inflation lasts long enough, it is
natural to consider the equilibrium solution of (\ref{FP}). Thus, we
have for the initial curvaton value $\sigma_{*}$
\beq
P_{\rm equil} = {1\over N}\exp{\left(-{8\pi^2\over
3H_{*}^4}V(\sigma_{*})\right)}
\eeq
where $H_{*}$ is the value of the expansion rate when our universe
crosses the horizon. As a consequence, the characteristic initial
value for the curvaton is given by the condition
\beq
\label{curvdistribution}
V(\sigma) \sim H^4/8\pi^2
\eeq
which is different from simply equating the classical force with the
quantum fluctuations $V'\sim H^3$ \cite{dynamics}. Thus during
inflation the subdominant curvaton has a probability distribution
that is fairly flat. In Sect.\ \ref{numerics}, where we present our
numerical results, this will not be considered as an additional
constraint although we display the limiting value
(\ref{curvdistribution}) in the Figures.

Given these constraints, for each point in the parameter space and
for each set of initial field values, we may then follow the
evolution of the field and the expansion rate to compute the
amplitude of the perturbation at the time when the curvaton finally
decays. We use of the $\Delta N$-formalism
\cite{gradient_expansion,deltaN,recent_deltaN}, where instead of
following the evolution of background fields and perturbations
around them, the evolution of superhorizon fluctuations is described
by considering an ensemble of homogenous patches of universe with
slightly different initial conditions. We compare the evolution of
two such copies: one has the initial field value $\sigma_*$, while
the other has
\begin{equation}
\sigma_+ = \sigma_* + \delta\sigma_{*}\ ,
\label{sigmapm}
\end{equation}
where $\delta\sigma_{*}={H_*}/{2\pi}$. Since we are interested in
the curvature perturbation on constant energy density hypersurfaces,
these two patches should then be evolved to a fixed value of $H$,
chosen so that for that value the curvaton has to a high accuracy
decayed completely in both patches. At this point the amount of
e-foldings in both patches is compared, and the difference, $\Delta
N$, gives the amplitude of curvature perturbation $\zeta$,
\begin{equation}
  \label{def_zeta}
  \zeta\equiv \Delta N= N(\sigma_+)-N(\sigma_*)
  =\Delta{\rm ln}\,a\left|\rule{0pt}{2.5ex}\right._{\rho}=({\rm ln}\,a)'
  \left|\rule{0pt}{2.5ex}\right._{\rho}\delta\sigma_{*}+\frac{1}{2}({\rm ln}\,a)''
  \left|\rule{0pt}{2.5ex}\right._{\rho}\delta\sigma_{*}^2
  +\ldots ~.
  \end{equation}
After the curvaton decay, the universe evolves adiabatically and the
curvature perturbation therefore remains constant $\dot{\zeta}=0$ on
superhorizon scales. The amplitude of large scale primordial
perturbations is fixed by CMB observations to be $\zeta\simeq 4.8
\times 10^{-5}$ \cite{cobenorm,wmap}.

\subsection{Qualitative features of evolution}
\label{qualitative_features}

Solving the full equations of motion analytically for arbitrary $n$
is not possible, although for the special case $n=0$ some useful
analytical approximations can be derived, as will be discussed in
Sect.\ \ref{samis}. Therefore, we will analyze the dynamics
numerically. We tackle the full numerics in Sect.\ \ref{numerics},
but let us first give a qualitative description of the dynamics
using na\"{\i}ve analytical estimates. The starting point here is
the observation \cite{turner} that for a scalar field $\sigma$
oscillating in a monomial potential $V \sim \sigma^{n+4}$ with
$n<6$, the average energy density obeys the scaling law
\begin{equation}
\rho_{\sigma} \propto a^{-6\frac{n+4}{n+6}} \ .
\label{scaling_law}
\end{equation}
Let us therefore consider the limit where either the quadratic term
or the interaction term $\sigma^{n+4}$ dominates the potential
(\ref{curvatonpot}). We find that there are three distinctive phases
in the evolution of the curvaton.

First, right after reheating, the curvaton is still effectively
massless with $V'' \ll H^2$ and the field value decreases very
slowly. Once $V'' \sim H^2$, the curvaton becomes massive and starts
to oscillate around the minimum of its potential. If the interaction
term dominates at this stage, the oscillations begin in the
$\sigma^{n+4}$ part of the potential and the curvaton energy density
scales as given by Eq.\ (\ref{scaling_law}). These oscillations
continue, until $\frac 12 m^2 \overline{\sigma}^2 \sim
\lambda\overline{\sigma}^{n+4}$, where $\overline{\sigma}$ is the
envelope of the oscillation. This marks the transition into curvaton
oscillations in the quadratic part of the potential, whence the
energy density scales as
\begin{equation}
\rho_{\sigma} \propto a^{-3} \, .
\label{quadrscaling}
\end{equation}
\begin{figure}
 \subfigure[Exact evolution.]{
  \includegraphics[width=5.5cm,angle=270]{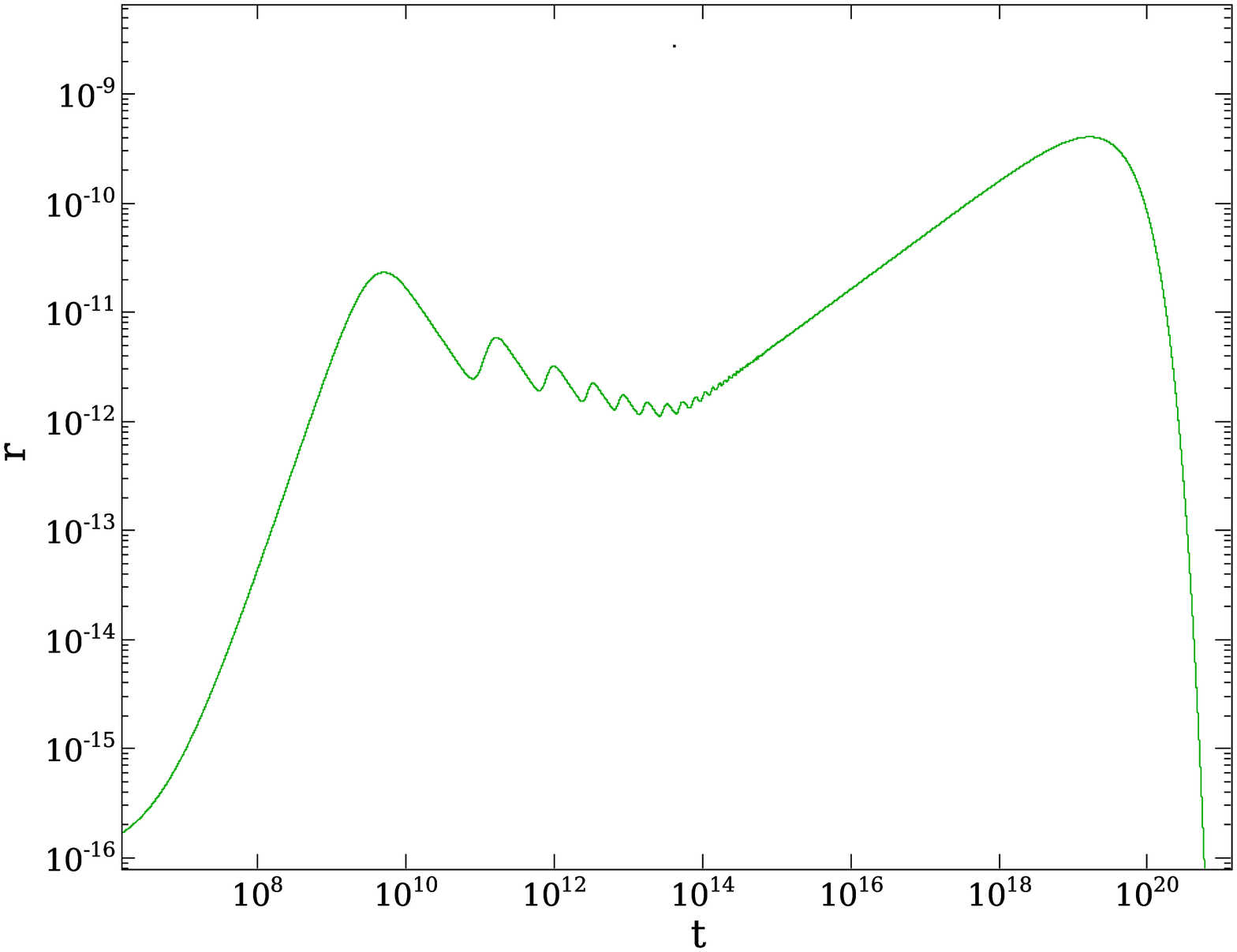}}
 \subfigure[Piecewise scaling-law approximation.]{
  \includegraphics[width=5.5cm,angle=270]{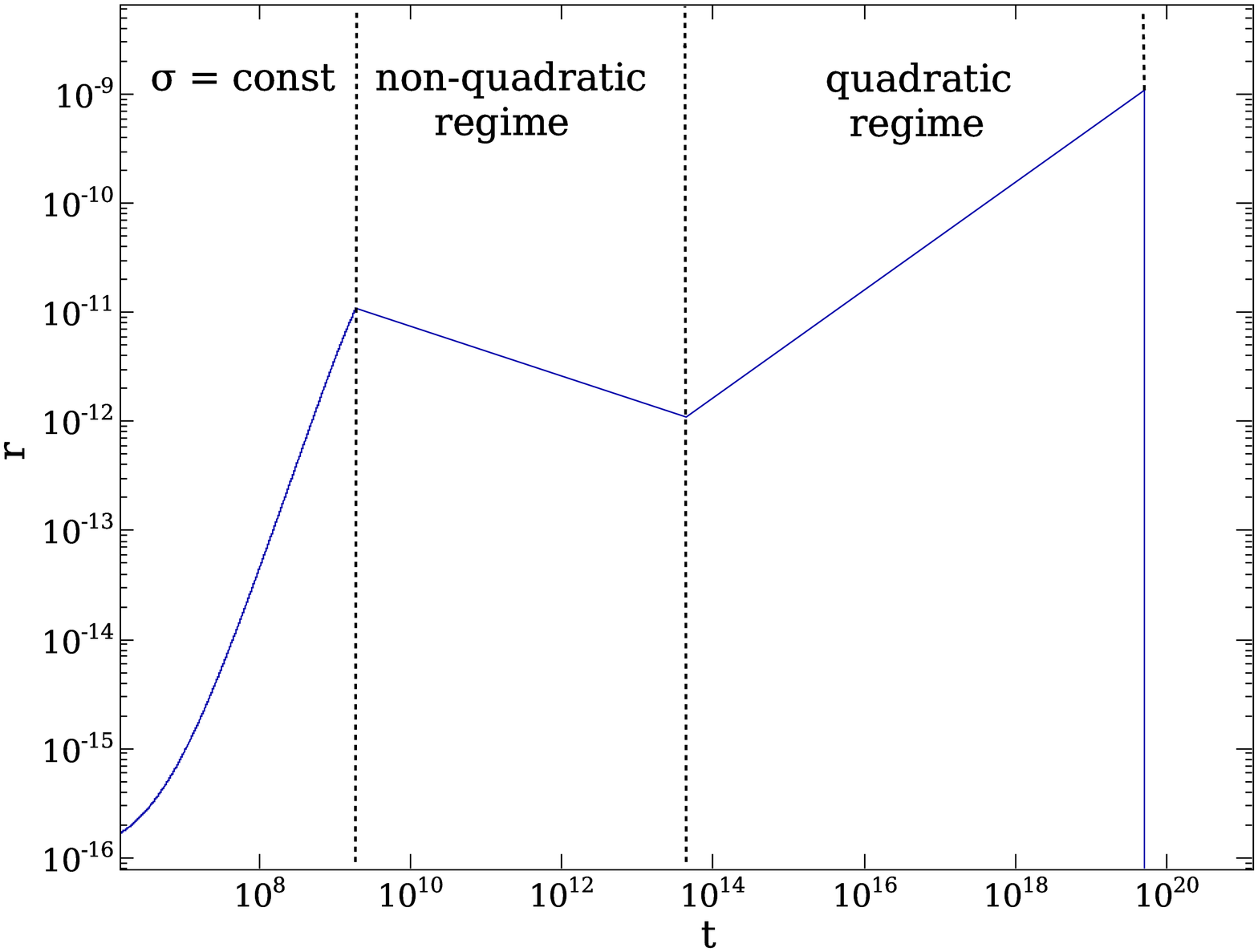}}
 \caption{Comparison between the exact evolution and scaling-law approximation of the relative curvaton energy density. Here the parameters have been chosen to be  $n = 2$, $m = 10^{-13}$ and $\Gamma = 10^{-20}$, and the inital conditions $r_* = 10^{-16}$ and $H_* = 10^{-7}$. The three different phases of curvaton evolution can be clearly seen: First the energy density stays constant and hence $r \propto a^{4}$, then it starts to oscillate in the non-quadratic part and hence $r\propto a^{-1/2}$ until it enters the regime of quadratic oscillations where $r \propto a$. Decay occurs when $\Gamma = H$.}
 \label{rhoevol}
\end{figure}
These three phases are clearly seen Fig.\ \ref{rhoevol},
which for a single choice of parameters displays the full numerical
solution of evolution of $r=\rho_ {\sigma}/\rho_{r}$, and then the
piecewise scaling-law approximation, where we assume that when a
single term dominates the potential, it scales according to Eq.\
(\ref{scaling_law}) until $H \sim \Gamma$, and the curvaton decays
instantaneously.

The more subdominant the curvaton is during its decay, the less
efficiently the initial perturbation in the curvaton field is
transmitted to metric fluctuations. However, since the curvaton is
very subdominant during inflation, i.e.\ its field value is small,
the quantum fluctuations produce relatively large perturbations when
compared to the field value. Since the perturbation itself can be
large, the curvaton does not need to be dominant at the decay.
Instead, it is sufficient that its relative density grows enough
from the given initial value. How much is "enough" will be discussed
in detail in Sect.\ \ref{numerics}.

\subsection{Breakdown of the scaling law and oscillation of the curvature perturbation}
\label{oscillations}

From the qualitative discussion above, one would infer that the
curvature perturbation varies smoothly and monotonously as the
parameter values are changed. However, this is not quite true as the
na\"{i}ve use of the scaling law (\ref{scaling_law}) does not
capture all the relevant dynamical details.

When using the scaling law (\ref{scaling_law}), one assumes the
validity of two approximations: first, that a single term in the
potential dominates completely; and second, that the oscillations
are rapid when compared to the Hubble time $H^{-1}$. Neither of
these assumptions is necessarily valid. The presence of two
different terms in the potential (\ref{curvatonpot}) could be taken
into account by generalizing the scaling law (\ref{scaling_law})
following the methods outlined in \cite{turner}. This would not
affect significantly the qualitative features of the results. The
breakdown of the latter assumption of rapid oscillations can however
have more significant effects. These can be seen in the numerical
solutions of the equations of motion which exhibit complicated
behaviour not qualitatively present in the simplified analytical
approximation. To demonstrate this, in Fig.\ \ref{fig:deltan} we plot
$\Delta N$ as a function of $H$ for fixed values of $H_*$, $r_*$ and
$\Gamma$ for two different curvaton mass values. For different
masses the moment of transition from the non-renormalizable part of
the potential to the quadratic part of the potential is different,
and this affects strongly the final value of $\Delta N$ which is
dictated by the duration of oscillations in the non-quadratic
regime.

\FIGURE[ht]{\label{fig:deltan}
\includegraphics[width=8cm,angle=270]{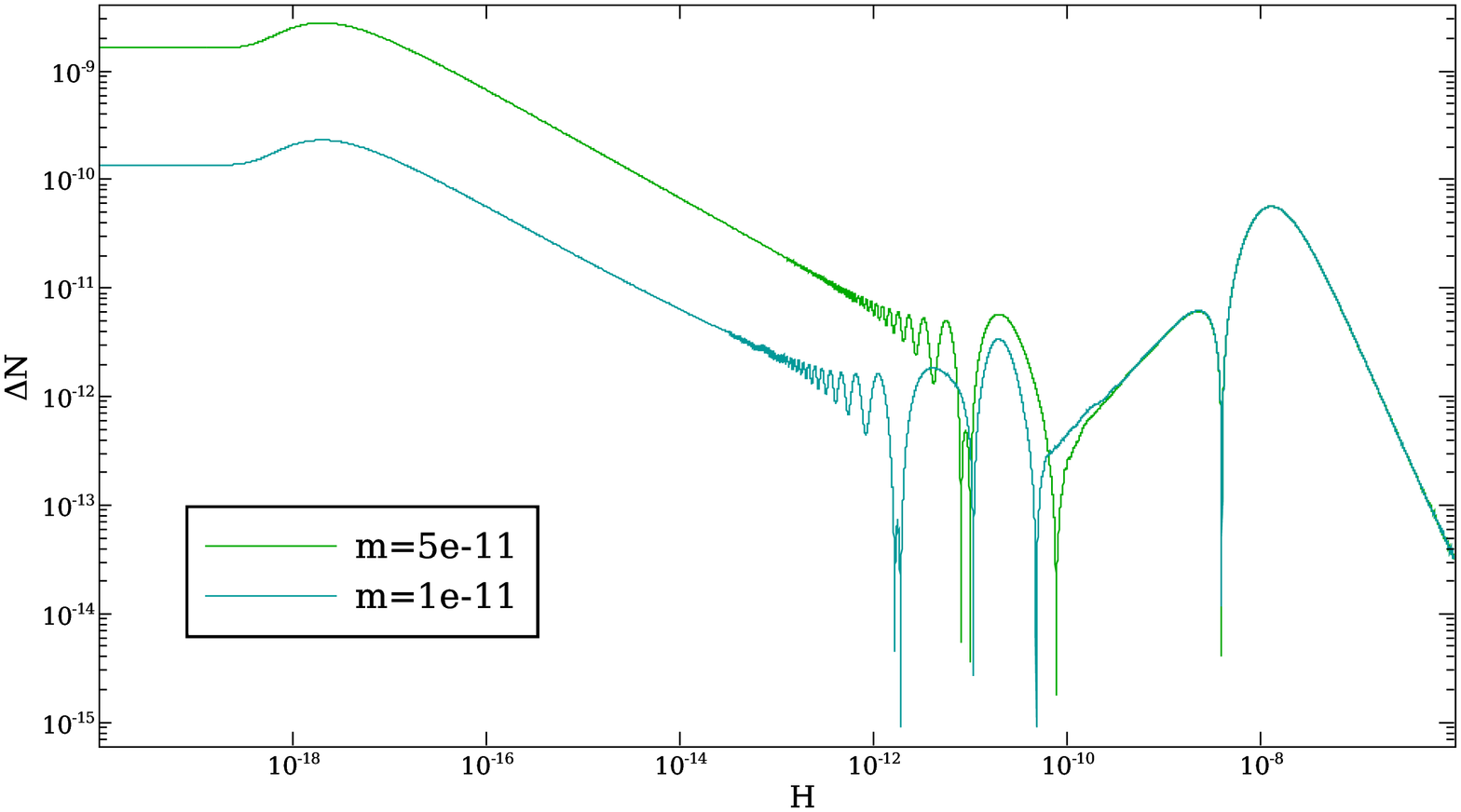}
\caption{$\Delta N$ as a function of $H$, (time
evolves from right to left). Both curves have $H_* = 10^{-6}$, $n =
4$, $\Gamma = 10^{-18}$ and $r_*=10^{-10}$, but masses are for the lower one $m =
10^{-11}$ and for the upper one $5 \cdot 10^{-11}$.
Here we demonstrate the non-trivial evolution of $\Delta N$.
Deep in the quadratic regime its oscillations become faster and faster
so that $\Delta N$'s evolution is given by a scaling law. However,
before the quadratic term in the potential starts to dominate, $\Delta N$ shows very complicated oscillatory behaviour.}}

From Fig.\ \ref{fig:deltan} it is clear that as the field value
oscillates in time, so does $\Delta N$. In the non-quadratic regime
$\Delta N$ oscillates with a large amplitude. If the transition to
the quadratic regime is slow compared to the oscillations in the
non-quadratic regime, the transition averages over several
oscillations. As a consequence, the final value of $\Delta N$ will
be a non-oscillatory function of the model parameters. However, if
the oscillation frequency in the non-quadratic potential is slow,
and the transition to the quadratic oscillations is rapid, then the
phase of the non-quadratic oscillation affects the final value of
$\Delta N$. If the parameters happen to be such that the transition
to the quadratic regime occurs at a maximum of the oscillation, a
relatively high value of $\Delta N$ freezes out. Similarly, if the
transition occurs at a minimum of the oscillation cycle, the final
value of $\Delta N$ will be much smaller. If the parameters
governing the moment of transition, such as the curvaton mass $m$,
are changed continuously, then the phase of the non-quadratic
oscillation during the transition also changes continuously. In the
space of the parameters this results in an oscillatory pattern in
$\Delta N$. This behaviour can be understood by observing that the
curvaton energy density at the beginning of oscillations can not be
expressed in terms of an amplitude of the envelope alone but also
depends on the phase of the oscillation, or equivalently on both the field $\sigma$ and its time derivative $\dot{\sigma}$, in a non-trivial way.
In effect, these act as two independent dynamical degrees of
freedom. If the transition from the interaction dominated part to
the quadratic region takes place at this stage, the initial
variation of the curvaton value $\sigma_{*}$ can therefore translate
in a non-trivial fashion into the final value of the curvature
perturbation.

In \cite{dynamics} it was shown that for a potential $V\sim
\sigma^{n+4}$ no oscillatory solutions exist if $n\geq6$. This means
that in the non-quadratic regime the curvaton merely decays and
hence no oscillations in $\Delta N$ occur. This is consistent with
the qualitative explanation of the oscillations discussed above.

\section{Numerical solutions}
\label{numerics}
%
\subsection{Solving the equations of motion}
%
We have explored the parameter space numerically for subdominant curvaton models that produce the curvature perturbation $\zeta\simeq 10^{-5}$ using the $\Delta N$ formalism with the patches defined as in Eq.\ (\ref{sigmapm}).

The solution of the full equations of motion (\ref{eom1})
oscillates more and more rapidly as it enters the regime dominated by the quadratic term in the potential. As each oscillation cycle requires several steps, solving the full equations of motion becomes increasingly slow as the rate of oscillations increase. To resolve this problem, we use the fact that deep in the quadratic regime the oscillations are very fast and hence the scaling law approximation Eq.\ (\ref{scaling_law}) becomes very good. Formally this corresponds to taking $\langle \dot{\sigma}^2 \rangle = \rho_\sigma$. Then we may switch from numerically evolving the equations of motion (\ref{eom1}) to solving the energy density evolution equations
\begin{eqnarray}
\label{edensityeqs}
& \dot{\rho_\sigma} = -\left(3H + \Gamma\right)\rho_\sigma &  \nonumber\\
& \dot{\rho_{\rm{r}}} = -4H\rho_{\rm{r}} + \Gamma \rho_\sigma &  \nonumber\\
& 3H^2 = \rho_{\rm{r}} + \rho_\sigma & \, .
\end{eqnarray}
Since the solution of these equations does not have oscillatory behaviour, the step size can be increased significantly without sacrificing accuracy. Thus we choose to evolve the full system until the quadratic term dominates the non-quadratic term by a very large factor, e.g.\ $10^5$ for $n \geqslant 2$. This value is chosen for convenience only and ideally should be as large as possible. This is the point where the numerical integration of the equations of motion becomes intolerably slow, and the smaller $n$ in Eq.\ (\ref{curvatonpot}) is, the slower the numerics becomes. With $n=0$, the system should be followed through a very large number of oscillation cycles. Thus for $n=0$ we have to stop intregration already when the quadratic-to-non-quadratic ratio is $10^2$ to prevent the time of integration to grow to several minutes.

At this point, we switch to the evolution equations for the energy densities (\ref{edensityeqs}) and follow the system until the curvaton has decayed, i.e.\ radiation
dominates the energy density of the curvaton by a very large factor, which we fix to be $10^{80}$. At this point we may compute the final value of $\Delta N$.

We are interested in how subdominant the curvaton is when it decays. This subdominance is described by the ratio of curvaton-to-radiation densities at the onset of decay, defined as
\begin{equation}
r_{\rm{decay}} = r(\Gamma\simeq H)~.
\label{rmax}
\end{equation}
Since the value of $r_{\rm{decay}}$ is not an explicit parameter in the model, we calculate it by
first solving the value for $\Gamma$ which gives the final amplitude of the perturbations to be the observed
magnitude, $10^{-5}$. $r_{\rm{decay}}$ is then calculated with these parameters. Hence the result of the
numerical simulations is a value for $r_{\rm{decay}}$ and $\Gamma$ for each parameter set $\{n, m, \lambda, H_*, r_*\}$.


\begin{figure}
\subfigure[$n=0$, $m=10^{-8}$]{
\includegraphics[width=8cm]{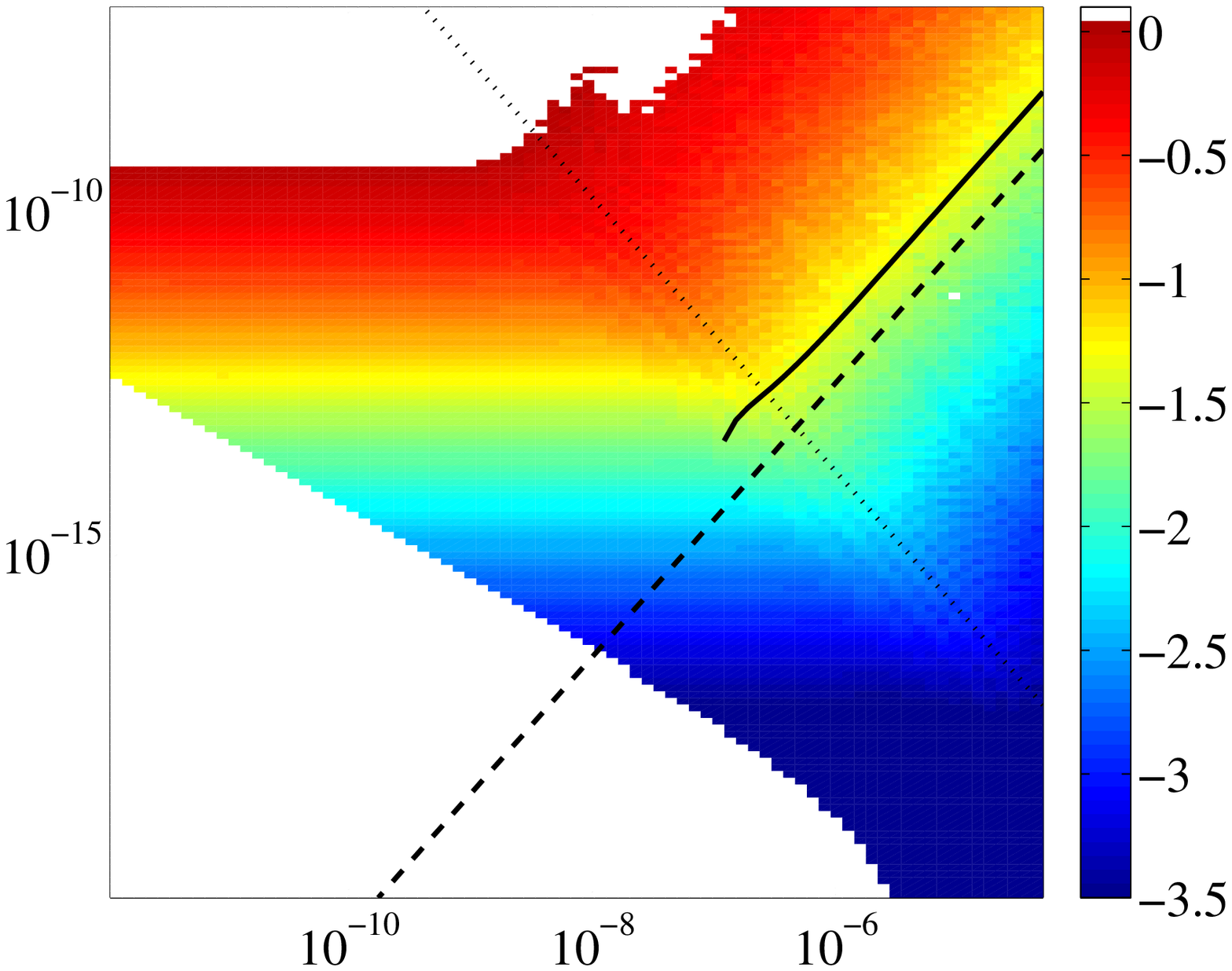}
}
\subfigure[$n=0$, $m=10^{-10}$]{
\includegraphics[width=8cm]{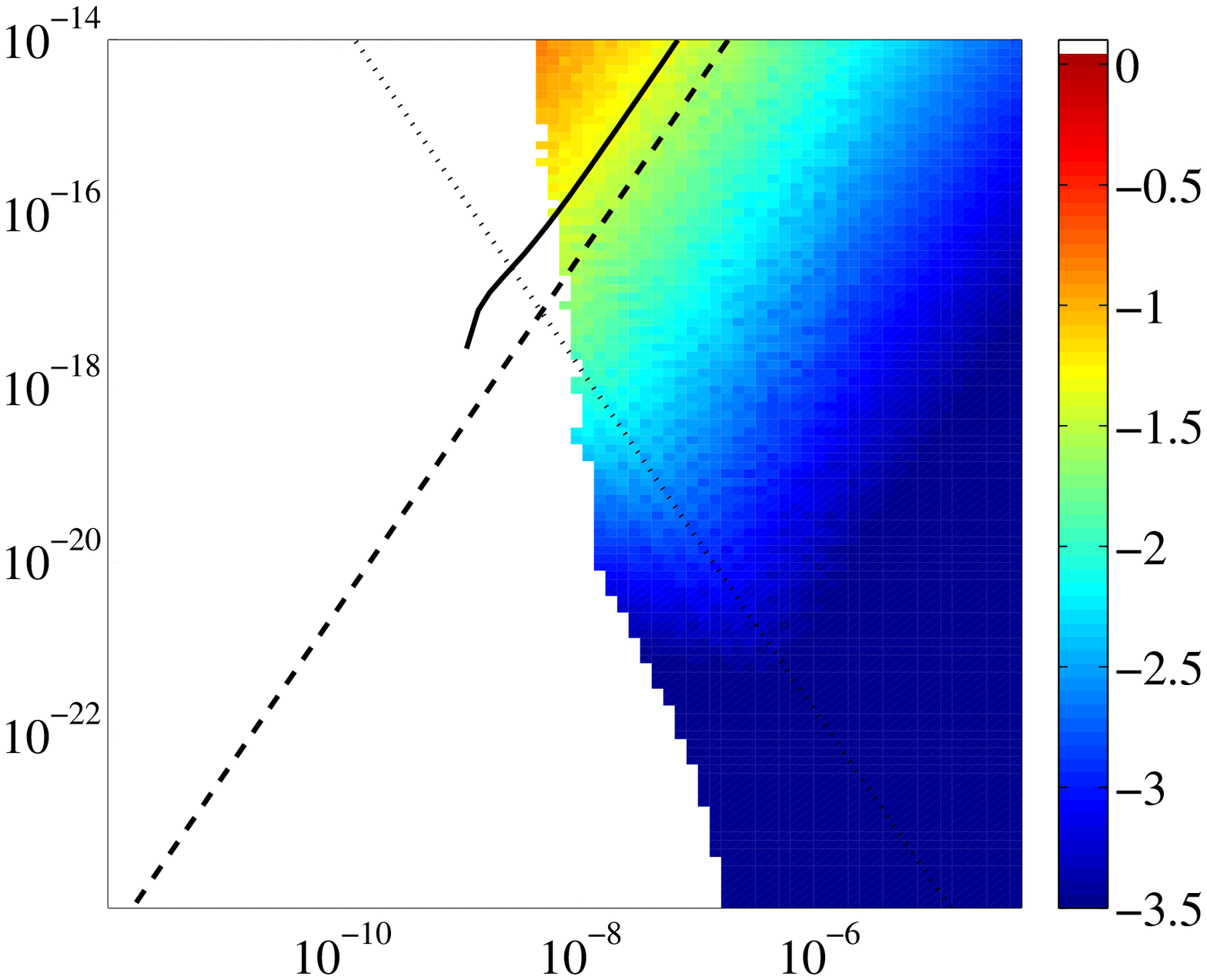}
}
\caption{$\log_{10} r_{\rm{decay}}$ as a function of $H_*$ (horizontal axis) and $r_*$ (vertical axis) for $n = 0$. The results are shown for a choice of coupling constant $\lambda = 10^{-7}$.
The white regions correspond to parameter values for which the amplitude of the curvature perturbation exceeds the observed value of $\zeta \sim 10^{-5}$, and are therefore immediately excluded.
The characteristic value of the initial curvaton probability distribution $V(\sigma) \sim H^4/8\pi^2$ is indicated by the dashed line with high probability below the line, while the area below the solid line corresponds to the condition for nearly scale invariant spectrum $V'' < 10^{-2}H_*^2$. Below the dotted line
non-quadratic terms in the potential are dynamically insignificant. }
\label{fig:contour1}
\end{figure}

\begin{figure}
\subfigure[$n=2$, $m=10^{-8}$]{
\includegraphics[width=8cm]{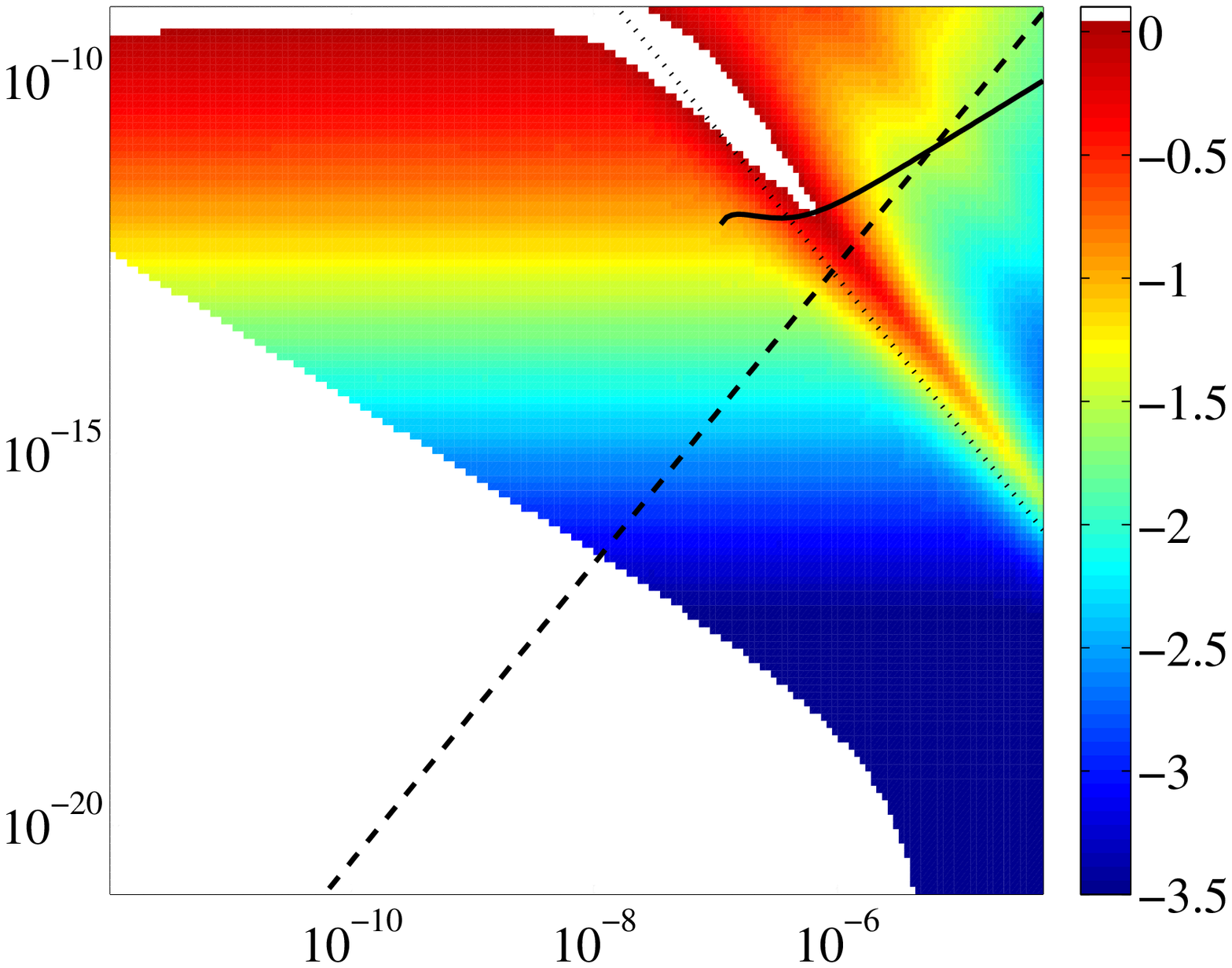}
}
\subfigure[$n=2$, $m=10^{-10}$]{
\includegraphics[width=8cm]{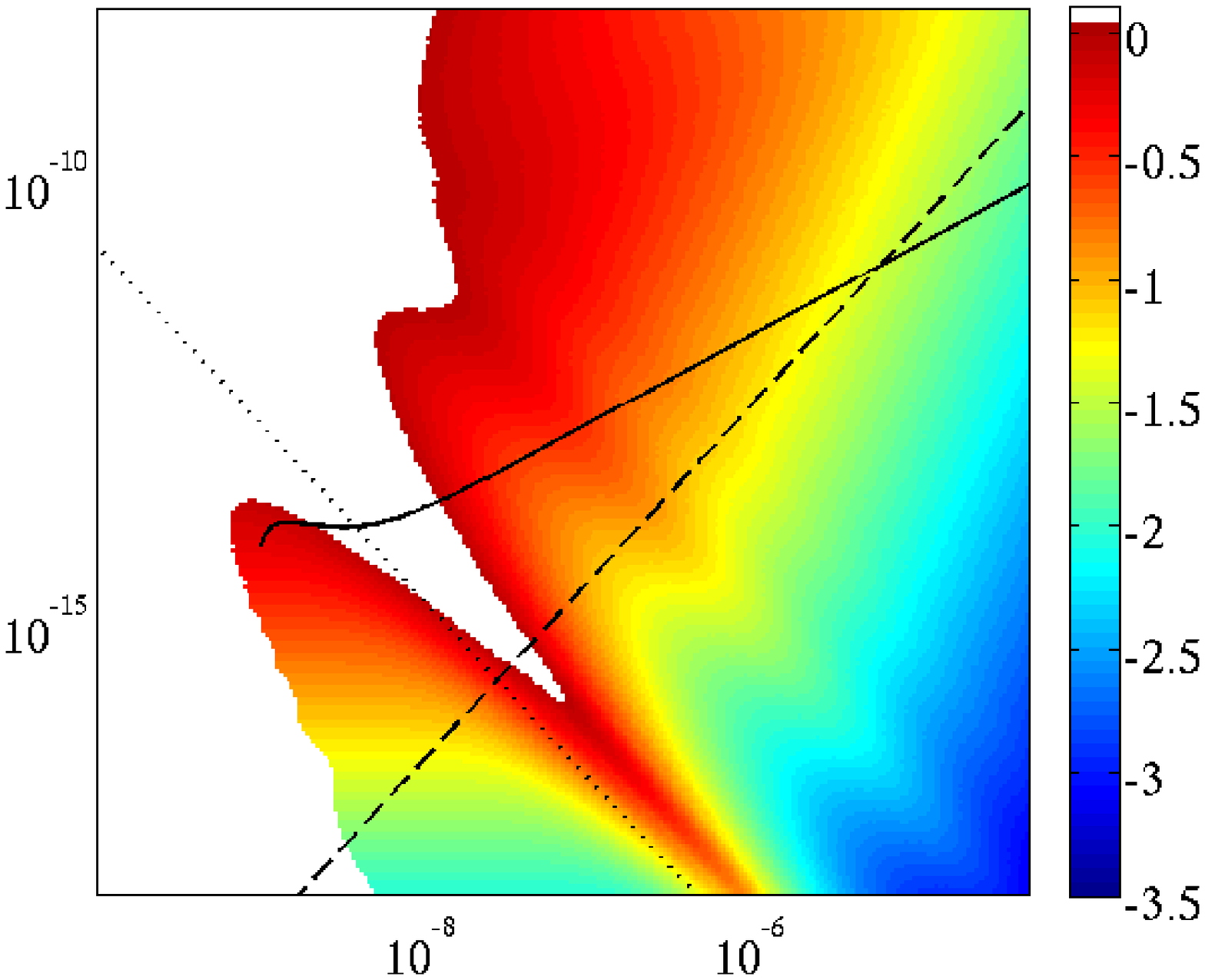}
}
\subfigure[$n=2$, $m=10^{-12}$]{
\includegraphics[width=8cm]{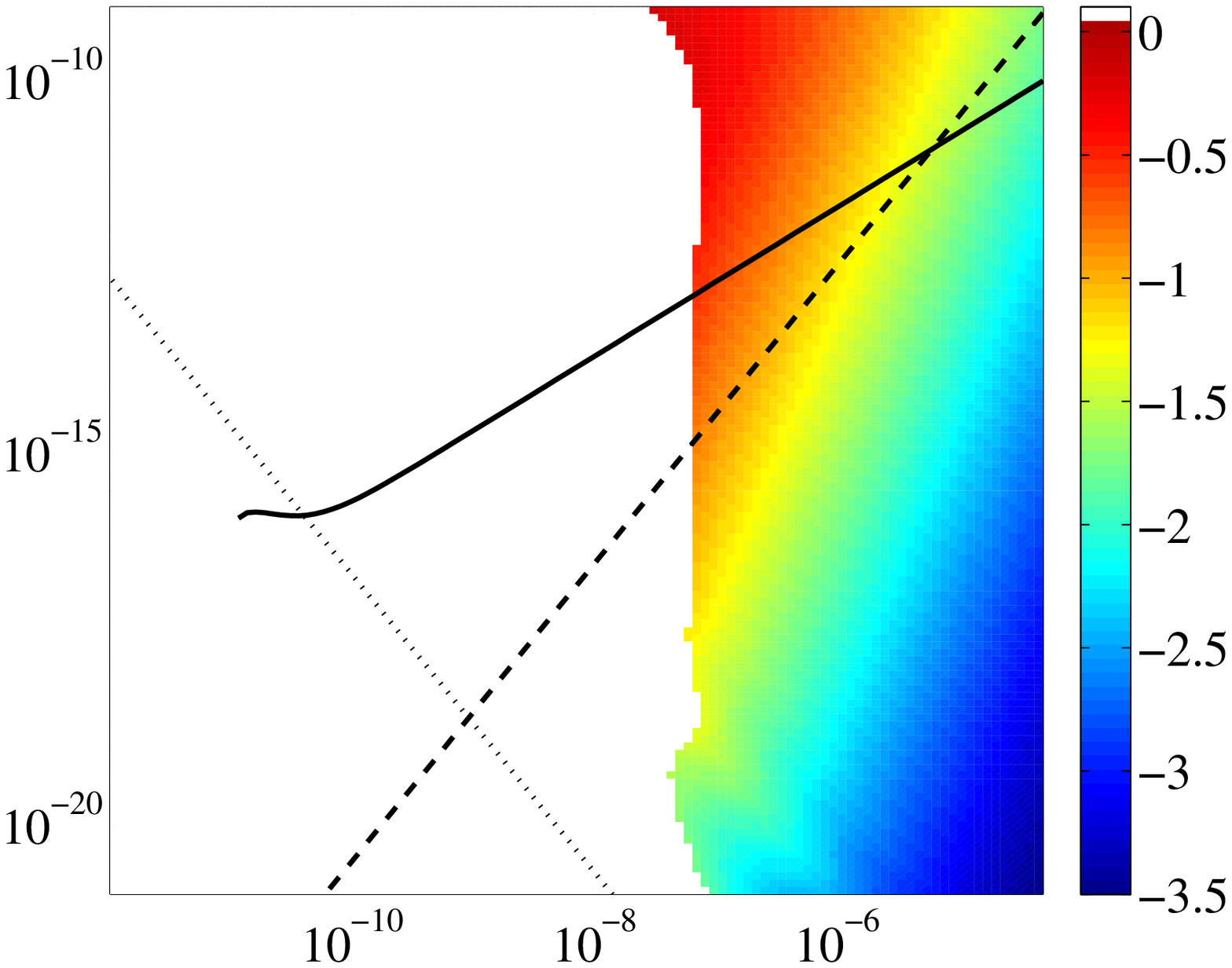}
}
\subfigure[$n=2$, $m=10^{-14}$]{
\includegraphics[width=8cm]{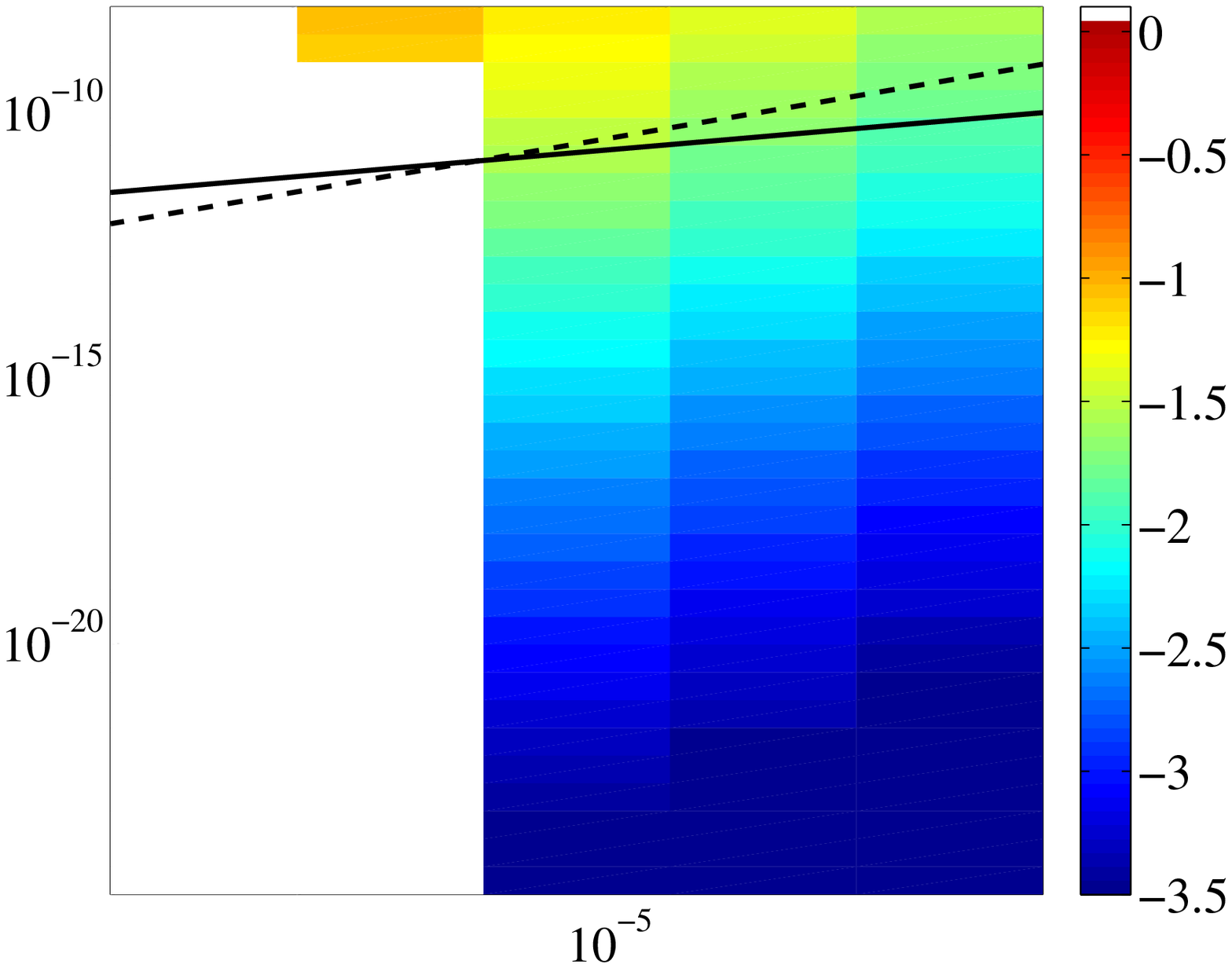}
}
\caption{$\log_{10} r_{\rm{decay}}$ as a function of $H_*$ (horizontal axis) and $r_*$ (vertical axis) for $n = 2$. The coupling constant is chosen to be $\lambda = 1$.
The white regions correspond to parameter values for which the amplitude of the curvature perturbation exceeds the observed value of $\zeta \sim 10^{-5}$, and are therefore immediately excluded.
The characteristic value of the initial curvaton probability distribution $V(\sigma) \sim H^4/8\pi^2$ is indicated by the dashed line with high probability below the line, while the area below the solid line corresponds to the condition for nearly scale invariant spectrum $V'' < 10^{-2}H_*^2$. Below the dotted line
non-quadratic terms in the potential are dynamically insignificant.}
\label{fig:contour2}
\end{figure}

\begin{figure}
\subfigure[$n=4$, $m=10^{-8}$]{
\includegraphics[width=8cm]{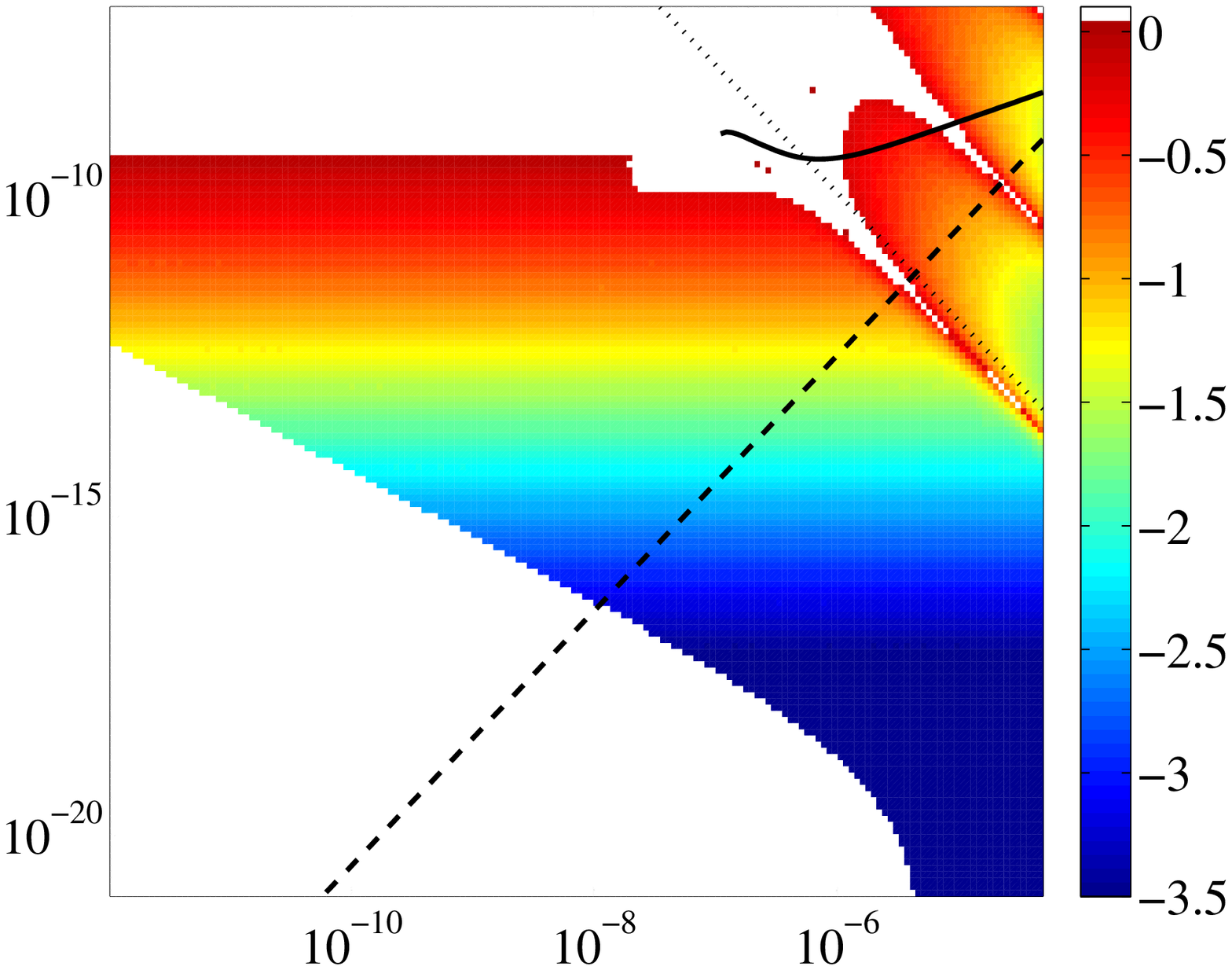}
}
\subfigure[$n=4$, $m=10^{-10}$]{
\includegraphics[width=8cm]{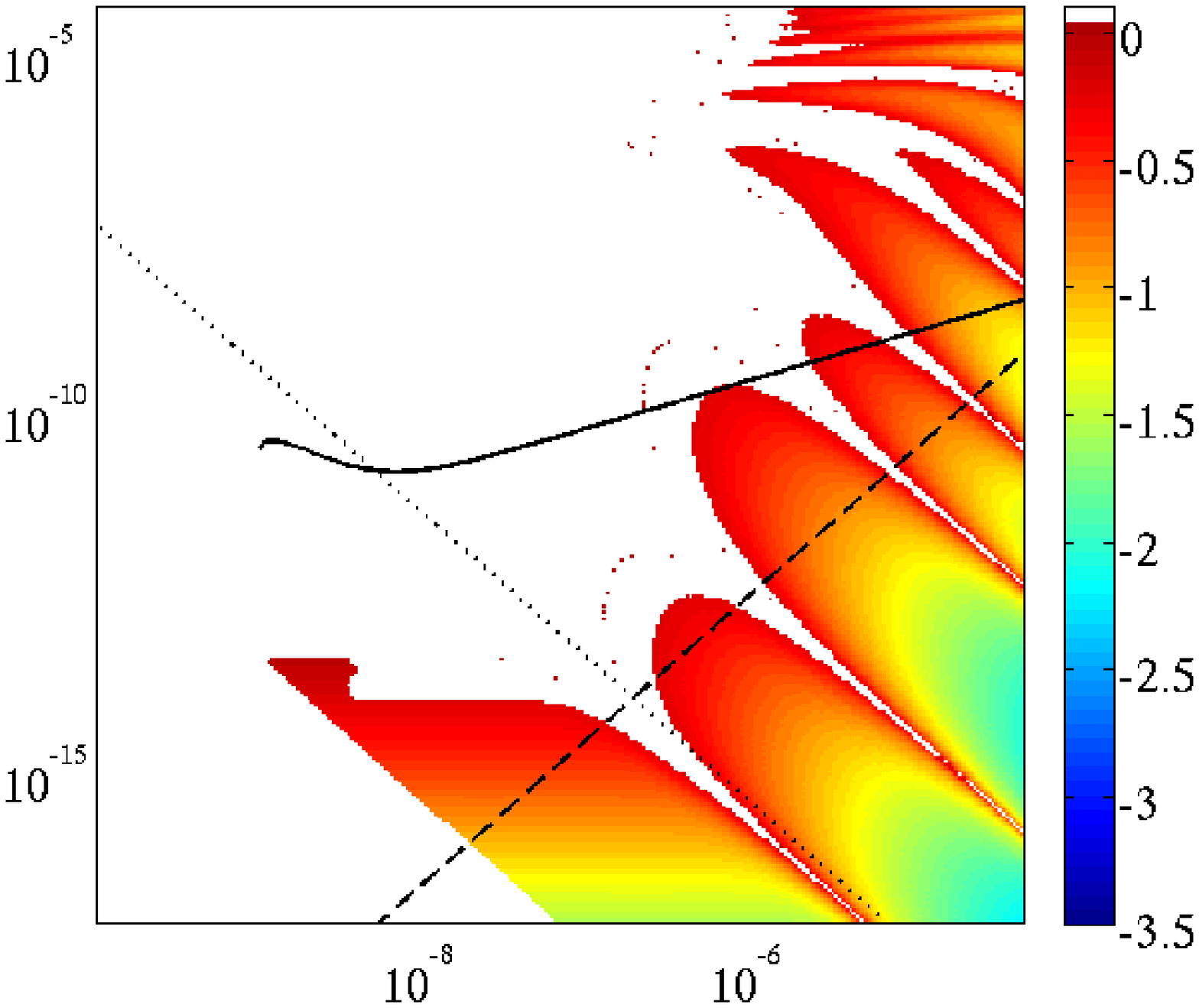}
}
\subfigure[$n=4$, $m=10^{-12}$]{
\includegraphics[width=8cm]{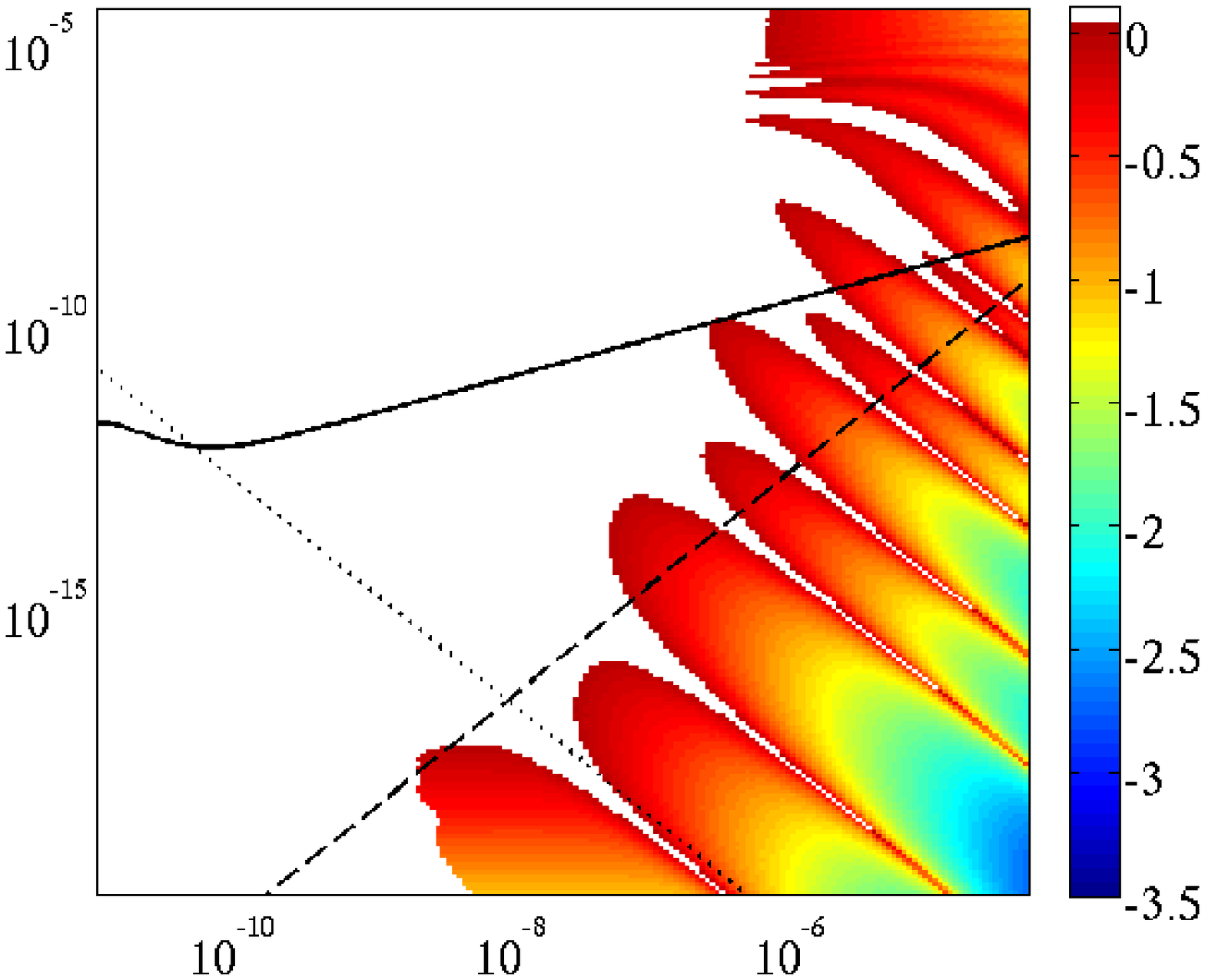}
}
\subfigure[$n=4$, $m=10^{-14}$]{
\includegraphics[width=8cm]{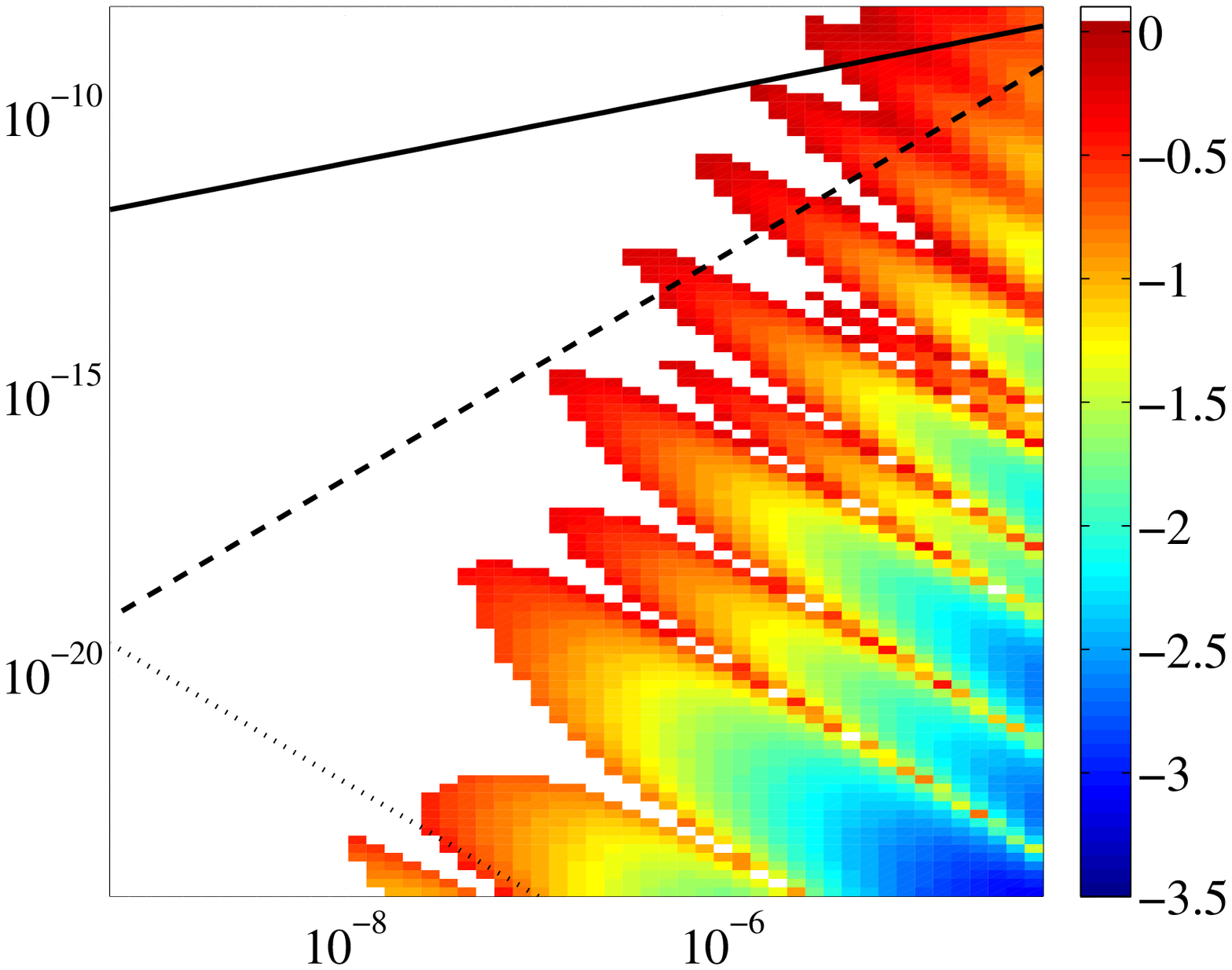}
}
\caption{Same as in fig \protect \ref{fig:contour2} but with $n=4$.}
\label{fig:contour3}
\end{figure}

\begin{figure}
\subfigure[$n=6$, $m=10^{-10}$]{
\includegraphics[width=8cm]{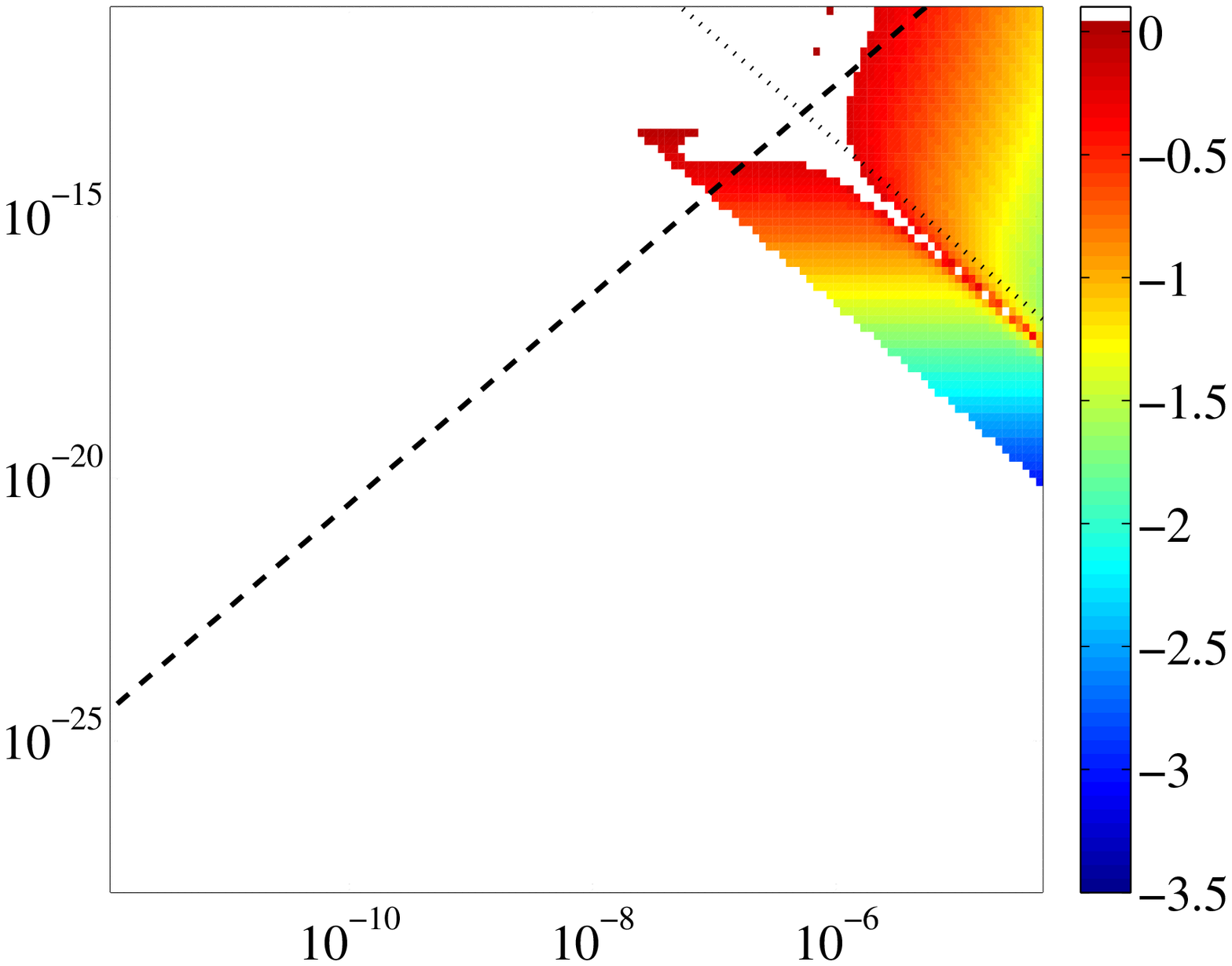}
}
\subfigure[$n=6$, $m=10^{-12}$]{
\includegraphics[width=8cm]{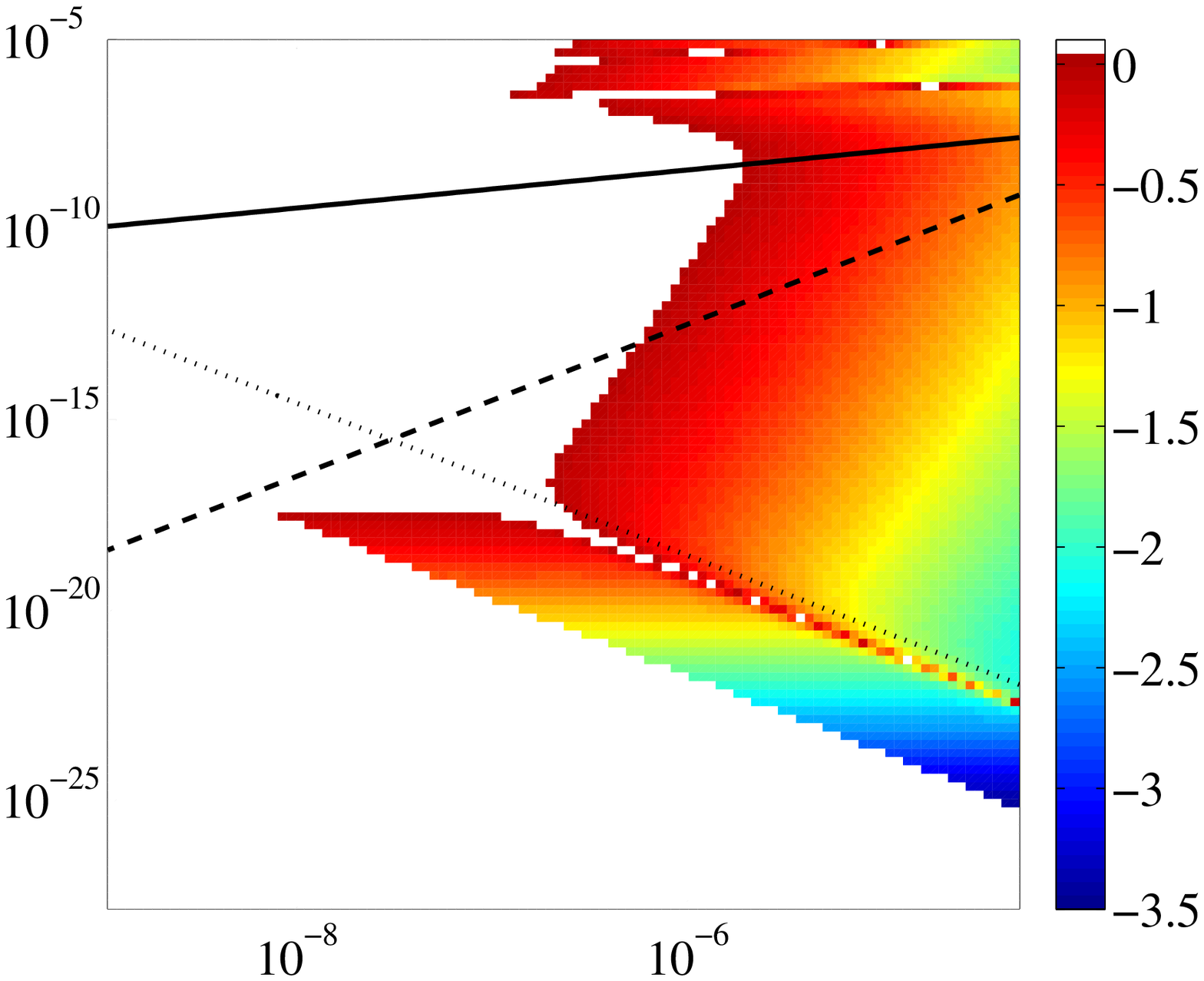}
}
\caption{Same as in fig \protect \ref{fig:contour2} but with $n=6$.}
\label{fig:contour4}
\end{figure}

\begin{figure}
\subfigure[$\log_{10} r_{\rm{decay}}$]{
\includegraphics[width=8cm]{n4_m12.eps}
}
\subfigure[$\log_{10} \Gamma$]{
\includegraphics[width=8cm]{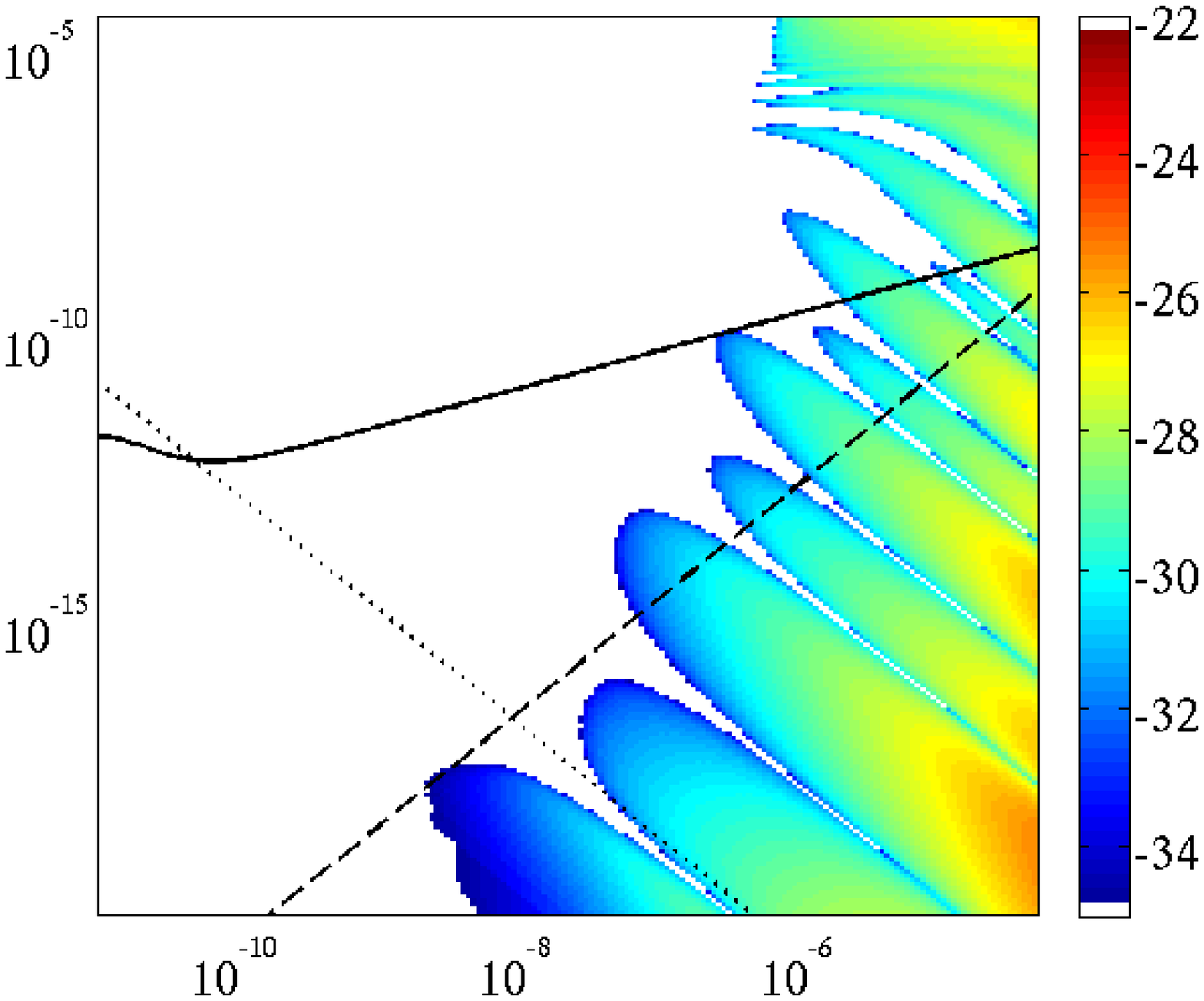}
}
\caption{$\log_{10} r_{\rm{decay}}$ and $\log_{10} \Gamma$ plotted as a function of $H_*$
(horizontal axis) and $r_*$ (vertical axis) for $m = 10^{-12}$ and $n=4$. Lines as in previous figures.}
\label{fig:gammacontour}
\end{figure}

\subsection{Perturbation amplitude in parameter space}

In Figs.\ \ref{fig:contour1} - \ref{fig:contour4} we have plotted
$\log_{10}r_{\rm{decay}}$ with different choices of $n$ and $m$ as a
function of $H_*$ and $r_*$. In addition to the value of
$r_{\rm{decay}}$ we have plotted also the constraint for near scale
invariance, $V'' = 10^{-2}H_*^2$ (solid line). The dotted line
corresponds to the equality of the quadratic and non-quadratic terms
in the potential during inflation; below the line the non-quadratic
term has no sizeable effect. We also show the position of the edge
of the curvaton probability distribution as given by Eq.\
(\ref{curvdistribution}) (dashed line); high probability is found
below this line.

For $n>0$ we set $\lambda$ to unity, i.e.\ the cut-off scale to the
Planck scale. We have, however, checked that decreasing $\lambda$
does not change the picture violently, but rather just moves the
features seen in Figs. \ref{fig:contour2} - \ref{fig:contour4}
slowly in the parameter space. For $n=0$, $\lambda$ is treated as a
free parameter up to the constraint $\lambda\ll 1$ coming from the
lightness of the curvaton in the interaction dominated regime.

The colder colors reflect a more subdominant curvaton and the warmer
more dominant. The white color indicates that no matter how dominant
the curvaton is, it cannot produce the observed amplitude of
perturbations. The dominant curvaton scenario corresponds to the
border between dark red and white, i.e.\ $r_{\rm{decay}} = 1$. The
numerical results demonstrate several features predicted by the
simplified qualitative description discussed in sections
\ref{qualitative_features} and \ref{oscillations}.

The decay time is constrained from below by the requirement of no isocurvature perturbations, as discussed in
Sect.\ \ref{modelconstr}. Adopting the limit on $\Gamma$ given in Eq. (\ref{decaybeforeDM}), some of the parameter space is ruled out; this is the source of
the cutoff at the lower left corner of the figures.

Oscillations of the value of $r_{\rm{decay}}$ in the parameter space can be clearly seen. As discussed before, their source is
the transition from the non-quadratic to the quadratic regime. Therefore,
as expected, below the dotted line in the figures, where the non-quadratic term never dominates, the oscillations of $r_{\rm{decay}}$ are absent.
Likewise, as there are no field oscillations in the $n=6$ potential, there are no ensuing oscillations of $r_{\rm{decay}}$ in the parameter space,
as can be seen in Fig.\ \ref{fig:contour4}. Furthermore, below the dotted line the value of $r_{\rm{decay}}$ depends only on $r_*$ and not on $H_*$. This is expected, since in the quadratic limit the relative curvaton perturbation is given by
\[ \frac{\delta \sigma_*}{\sigma_*} = \frac{H_*/ 2\pi}{\sqrt{6r_*}H_*/m} =
\frac{m}{2\pi\sqrt{6r_*}} \, .\]

Some of the parameter space of different curvaton models, as
specified by $n$, and $m$, and by the inital conditions $r_*$ and
$H_*$, is clearly ruled out. This is the white area in the plots.
Large parts of the parameter space are however allowed, and for
these areas, we have calculated the degree of subdominance at the
time of curvaton decay. The figures clearly demonstrate that a
curvaton model is feasible even when it is subdominant by a factor
of $10^{-3}$ during decay. However, one should note that a very
subdominant curvaton usually implies a large inflationary scale, as
is also evident in the figures. This might not be a problem but as
the inflationary scale grows the curvature perturbations generated
during inflation also grow and can become significant \cite{mixed}
which needs to be taken into account in the analysis. For single
field slow roll inflation, the inflaton perturbations are negligible
if $H_{*}\ll 10^{-5}\sqrt{\epsilon_{*}}$.

The results given by the numerical code described above were also
checked with an independently developed program, to verify that the
features seen are not numerical errors.

%
%
%
%
\section{Analytical results for $n=0$}
\label{samis}
%
%
%
%
\subsection{Solving the equation of motion}
\label{samis}
%

The numerical results discussed in the previous section can be
contrasted with an analytical estimate in the special case of $n=0$
for which we the potential (\ref{curvatonpot}) reads
\beq
\label{sigma4pot}
V=\frac{1}{2}m^2\sigma^2+\lambda\sigma^4\ .
\eeq
For this particular potential, it is possible to find a reasonably
accurate analytical approximation for the exact solution of curvaton
equation of motion. To some extent, the analysis serves as a useful consistency
check of our numerical results, but can also be of certain interest
of its own as it allows in principle to treat arbitrary
values of the coupling $\lambda$ analytically.

In a radiation dominated background the equation of motion for a
homogeneous curvaton with the potential (\ref{sigma4pot}) can be
written as
  \beq
  \label{f_eom_exact}
  \frac{{\rm d}^2f(a)}{{\rm d}a^2}+c_1f(a)^3+c_2 a^2f(a)=0
  \eeq
where $a=(t/t_{*})^{1/2}$ is the scale factor, we have defined
$f=a\sigma/\sigma_{*}$ and the star denotes the end of inflation.
The constants are defined as $c_1=4\lambda\sigma_{*}^2/H_{*}^2$ and
$c_2=m^2/H_{*}^2$. To identify $\sigma_{*}$ with the curvaton value
at $t_{*}$, we should set the initial condition
$\sigma(t_{*})=\sigma_{*}$. However, as the field is light at early
times and evolves slowly, we choose instead to formally set the
initial conditions $\sigma(t\rightarrow 0)=\sigma_{*}$ and
$\dot{\sigma}(t\rightarrow 0)=0$ in the asymptotic past. This
corresponds to
  \beq
  \label{initial_conditions}
  f(a)|_{a=0}=0\ ,~~f'(a)|_{a=0}=1\ ,
  \eeq
and yields $\sigma(t_{*})=\sigma_{*}(1-{\cal O}(c_1))\simeq
\sigma_{*}$, assuming $c_1>c_2$.

Exact solutions for Eq. (\ref{f_eom_exact}) can be found if one of
the constants $c_{1}$ and $c_{2}$ vanishes. In the absence of the
quadratic term, $c_2=0$, the solution obeying the initial conditions
(\ref{initial_conditions}) is given by the elliptic sine
$f_{m=0}(a)=2^{1/4}c_1^{-1/4} {\rm sn}(2^{-1/4} c_1^{1/4} a,-1)$.
This can be well approximated by the leading term of its
trigonometric series
\beq
  \label{f_trigon}
f_{m=0}(a)\simeq 2^{1/4}c_1^{-1/4} {\rm
sin}(b\, 2^{-1/4} c_1^{1/4} a)
\eeq
with $b=\pi (2 K(-1))^{-1}\simeq 1.1981$. The approximative solution
obviously satisfies a differential equation obtained by setting
$m=0$ in Eq. (\ref{f_eom_exact}) and making the replacement
$f(a)^3\rightarrow b^2c_1^{1/2}2^{-1/2} f(a)$.

\begin{figure}[!h]
 \centering
 \includegraphics[width=15cm, height=4 cm]{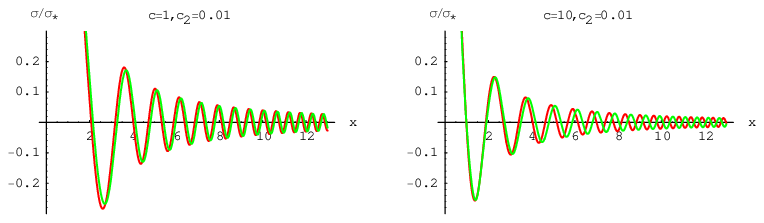}
 \caption{Comparison of the solutions for the curvaton field $\sigma = \sigma_{*} f(a)/a$
  obtained from the exact equation (\protect \ref{f_eom_exact}) (green) and the approximative one
  (\protect \ref{f_eom}) (red).}
 \label{fig:sami1}
 \end{figure}

This observation motivates us to study a much simpler linear
equation
  \beq
  \label{f_eom}
  \frac{{\rm d}^2f(x)}{{\rm d}x^2}+\left(c+x^2\right)f(x)=0\ ,
  \eeq
instead of Eq. (\ref{f_eom_exact}). Here a yet another redefinition of
variables has been made
  \beq
  \label{x_ja_c}
  x=a c_2^{1/4}\ ,
  ~~~~~~c=\frac{b^2\sqrt{2\lambda}\,\sigma_{*}}{m}\ .
  \eeq
Instead of the initial conditions (\ref{initial_conditions}) we need
to impose
  \beq
  \label{initial_conditions_app}
  f(x)|_{x=0}=0\ ,~~f'(x)|_{x=0}=b c_2^{-1/4}\ ,
  \eeq
to match with the desired exact solution of (\ref{f_eom_exact}) in
the limit $m\rightarrow 0$. By choosing the initial conditions
(\ref{initial_conditions_app}) we implicitly restrict our analysis
on the parameter values for which the quartic term is dominant or
nearly dominant in (\ref{f_eom_exact}) at early times, $c_1\gtrsim
c_2$. Since Eq. (\ref{f_eom}) yields the correct approximative
solution in the quartic limit $m\rightarrow 0$ and reduces to Eq.
(\ref{f_eom_exact}) at late times $x\gg 1$ when the quadratic term
dominates the dynamics, we may expect it to yield a reasonable
estimate for the asymptotic solutions of (\ref{f_eom_exact}).
Comparison with numerical results confirms that indeed this is the
case, as illustrated in Fig. \ref{fig:sami1}.
As can be seen, the solutions of Eqs. (\ref{f_eom}) and
 (\ref{f_eom_exact}) differ by a slowly varying phase but the
amplitudes agree quite well. This seems to be a generic outcome and
not just a coincidence for the parameter values shown in Fig.
\ref{fig:sami1}. For our current purposes the phase difference is
irrelevant as it would affect the curvature perturbation at late
times $mt \gg 1$ only through terms suppressed by $m/H$ or $r$.

\subsection{Asymptotic solution}

\begin{figure}[!h]
 \centering
 \includegraphics[width=15cm, height=4 cm]{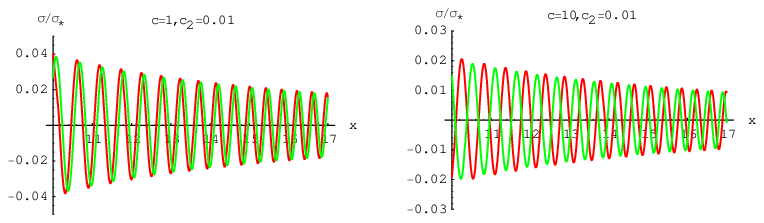}
 \caption{Comparison of the exact solution (green) for the curvaton field $\sigma = \sigma_{*}
 f(a)/a$ obtained from Eq.\ (\protect \ref{f_eom_exact}) and the asymptotic result (red)
 from Eq.\ (\protect \ref{sigma_as}).}
  \label{fig:sami2}
 \end{figure}

The general solution for (\ref{f_eom}) can be expressed in terms of
parabolic cylinder functions which admit simple asymptotic
representations in the limit of large argument \cite{GR}. Focusing
on late times $x=\sqrt{2 m t}\gg c,~x\gg 1$, Eq. (\ref{f_eom})
yields the asymptotic result
    \beq
  \label{sigma_as}
  \sigma_{\rm as}(t)\sim\frac{\sigma_{\rm osc}}{(mt)^{\frac{3}{4}}}\,{\rm
  sin}\left(mt+\frac{\pi}{8}+\delta\right)\ ,
  \eeq
where
  \baq
  \label{sigma_osc}
  \sigma_{\rm osc}&=&\sigma_{*}\frac{b\sqrt{\pi}e^{-\pi
  c/8}2^{-3/4}}{|\Gamma\left(\frac{3}{4}+i\frac{c}{4}\right)|}\
  ,
  \\
  \label{delta}
  \delta&=&\frac{c}{4}\,{\rm ln}(2 m t
  )-{\arg}\left(\Gamma\left(\frac{3}{4}+i\frac{c}{4}\right)\right)\
  ,
  \eaq
and the initial conditions (\ref{initial_conditions_app}) have been
used. Fig. \ref{fig:sami2} shows a comparison between Eq. (\ref{sigma_as}) and the
exact solutions of (\ref{f_eom_exact}) for the same parameter
choices as in Fig. \ref{fig:sami1} but for later time events $x\gg c$ where
the asymptotic solution is expected to be valid.
Again it can be observed that the results are offset by a slowly
varying phase difference, which manifests itself as the first term
in Eq. (\ref{delta}), but the time evolution of the amplitude is well
described by the asymptotic solution Eq. (\ref{sigma_as}).

\begin{figure}[!h]
 \centering
 \includegraphics[width=6cm, height=4 cm]{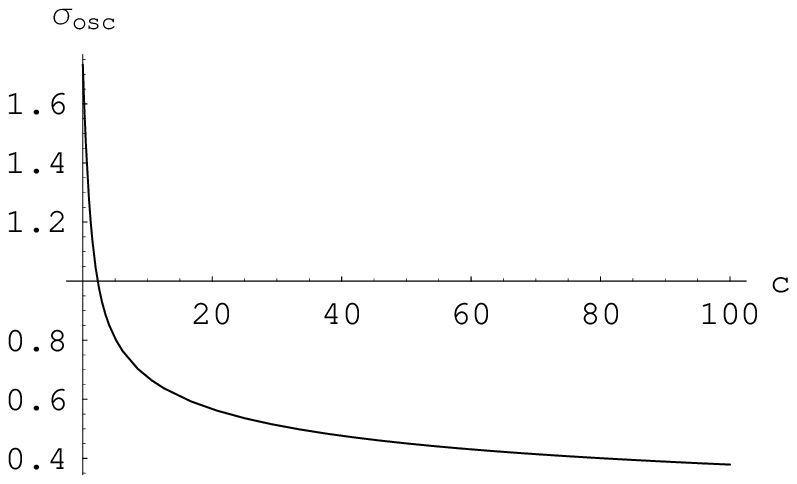}
 \caption{The effective curvaton amplitude (\protect \ref{sigma_osc}) $\sigma_{\rm osc}$ at the beginning of
 quadratic oscillations, $c\sim 2\sqrt{\lambda}\sigma_*/m$ measures the relative strength of the interaction term
 at the end of inflation.}
  \label{fig:sami3}
 \end{figure}

The dependence of $\sigma_{\rm osc}$ on the magnitude of the quartic
term in Eq. (\ref{sigma4pot}) is illustrated in Fig \ref{fig:sami3}.
For the interaction dominated case $c\gg 1$, Eq. (\ref{sigma_osc})
yields $\sigma_{\rm osc}\simeq \sigma_{*} c^{-1/4} 2^{-3/4}b$ which
according to (\ref{sigma_as}) leads to an estimate $\bar{\sigma}\sim
\sigma_{*}c^{-1/4}(H/m)^{3/4}$ for the envelope of oscillations in
the quadratic regime. This is in agreement with the result obtained
using the simple scaling law (\ref{scaling_law}) and choosing
$4\lambda\bar{\sigma}^2=m^2$ as the transition between quartic and
quadratic regimes. In the limit $c\rightarrow 0$ where the quadratic
term dominates throughout the evolution, $\sigma_{\rm
osc}\rightarrow \sigma_{*} b\sqrt{\pi}2^{-3/4}/\Gamma(3/4)$ and Eq.
(\ref{sigma_as}) reduces to an asymptotic representation of a Bessel
function. Up to the constant $b\simeq 1.1981$ related to the choice
of initial conditions (\ref{initial_conditions_app}) engineered for
the case $c\gtrsim 1$, this coincides with the exact solution for
the quadratic case obeying the initial conditions
(\ref{initial_conditions}).

%
\subsection{Curvature perturbation}
%
%
Using the result (\ref{sigma_as}), the curvaton energy density at
late times $mt\gg 1,~mt\gg c^2$ can be written as
  \beq
  \label{rho_sigma_as}
  \rho_{\sigma}\simeq\frac{m^2\sigma_{\rm osc}^2}{2(mt)^{3/2}}\left(1+{\cal {O}}(H/m)\right)\ .
  \eeq
The contribution from the quartic part of the potential
(\ref{sigma4pot}) is negligible as equations (\ref{sigma_as}) and
(\ref{sigma_osc}) yield $\lambda\sigma^2/m^2\lesssim
c^2/(mt)^{3/2}$. Neglecting the non-gravitational interactions
between the curvaton and the dominant radiation component, the total
energy density reads
  \beq
  \label{rho_tot}
  \rho=\rho_{r*} a^{-4}+\rho_{\sigma}\ ,
  \eeq
where $\rho_{r*}\simeq 3 H_{*}^2$ is the radiation energy density at
the initial time $t_{*}$.

The curvature perturbation (\ref{def_zeta}) at some time $t$ can be
straightforwardly computed by varying (\ref{rho_tot}) with respect
to the initial curvaton value $\sigma_{*}$ and keeping the final
energy density $\rho$ at $t$ fixed. To leading order in
$r=\rho_{\sigma}/\rho_r$, this leads to the result
  \beq
  \label{zeta_exp}
  \zeta\simeq\frac{r}{2}\frac{\sigma_{\rm osc}'}{\sigma_{\rm
  osc}}\delta\sigma_{*}+\frac{r}{4}\left(\frac{\sigma_{\rm osc}''}{\sigma_{\rm osc}}
  +\left(\frac{\sigma_{\rm osc}'}{\sigma_{\rm
  osc}}\right)^2\right)\delta\sigma_{*}^2+\ldots=
  \sum_{n=1}^{\infty}\frac{\alpha_n}{n!}\delta\sigma_{*}^n\  ,
  \eeq
where
  \beq
  \label{alpha_n}
  \alpha_{n}=\frac{r}{4\sigma_{\rm osc}^2}\sum_{m=0}^{n}\frac{n!}{m!(n-m)!}
  \frac{\partial^{n-m}\sigma_{\rm osc}}{\partial\sigma_{*}^{n-m}}
  \frac{\partial^{m}\sigma_{\rm osc}}{\partial\sigma_{*}^{m}}\ ,
  \eeq
and
  \beq
  \label{r_t}
  r\simeq \frac{\sqrt{2}}{3}\,\sigma_{\rm
  osc}^2\left(\frac{m}{H}\right)^{1/2}\ .
  \eeq
The prime in (\ref{zeta_exp}) denotes differentiation with respect
to $\sigma_{*}$ and the amplitude of perturbations is estimated by
$\delta\sigma_{*}\sim H_{*}/(2\pi)$. In the sudden decay
approximation, the curvaton is assumed to decay instantaneously at
the time $H=\Gamma$ where, up to factors of unity, $\Gamma$ is given
by the model dependent perturbative curvaton decay rate. The final
value of the curvature perturbation under this assumption is
obtained by setting $H=\Gamma$ in (\ref{zeta_exp}). Here we assume
that the universe evolves adiabatically after the curvaton decay and
hence $\dot{\zeta}=0$.

It is also straightforward to derive a formally non-perturbative
expression for the curvature perturbation (see also
\cite{many_curvatons}) which can be directly compared with our
numerical results. By performing a finite shift
$\sigma_{*}\rightarrow \sigma_{*}+\Delta\sigma_{*}$ in
(\ref{rho_tot}) but keeping the energy density $\rho$ fixed, we
arrive at a fourth order equation for the curvature perturbation
$\zeta={\rm ln}\,(a+\Delta a)-{\rm ln}\,a$
  \beq
  \label{zeta_nb}
  (1+r(\sigma_{*}))\,e^{4 \zeta}-\frac{r(\sigma_{*}) \sigma_{\rm
  osc}^2(\sigma_{*}+\Delta\sigma_{*})}{\sigma_{\rm
  osc}^2(\sigma_{*})}\,e^{\zeta}-1=0\ .
  \eeq

\subsection{Comparing with numerical estimates}

\label{n_0_lambda}

Equation (\ref{zeta_nb} ) can be readily solved using standard methods. Qualitative
features of the solutions are illustrated in Fig. \ref{fig:vertailu}
below, which shows $\zeta(\sigma_{*})$ for three different choices
of the coupling $\lambda$ together with the corresponding numerical
results.
 \begin{figure}[!h]
 \centering
 \includegraphics[scale=0.5,angle=-90]{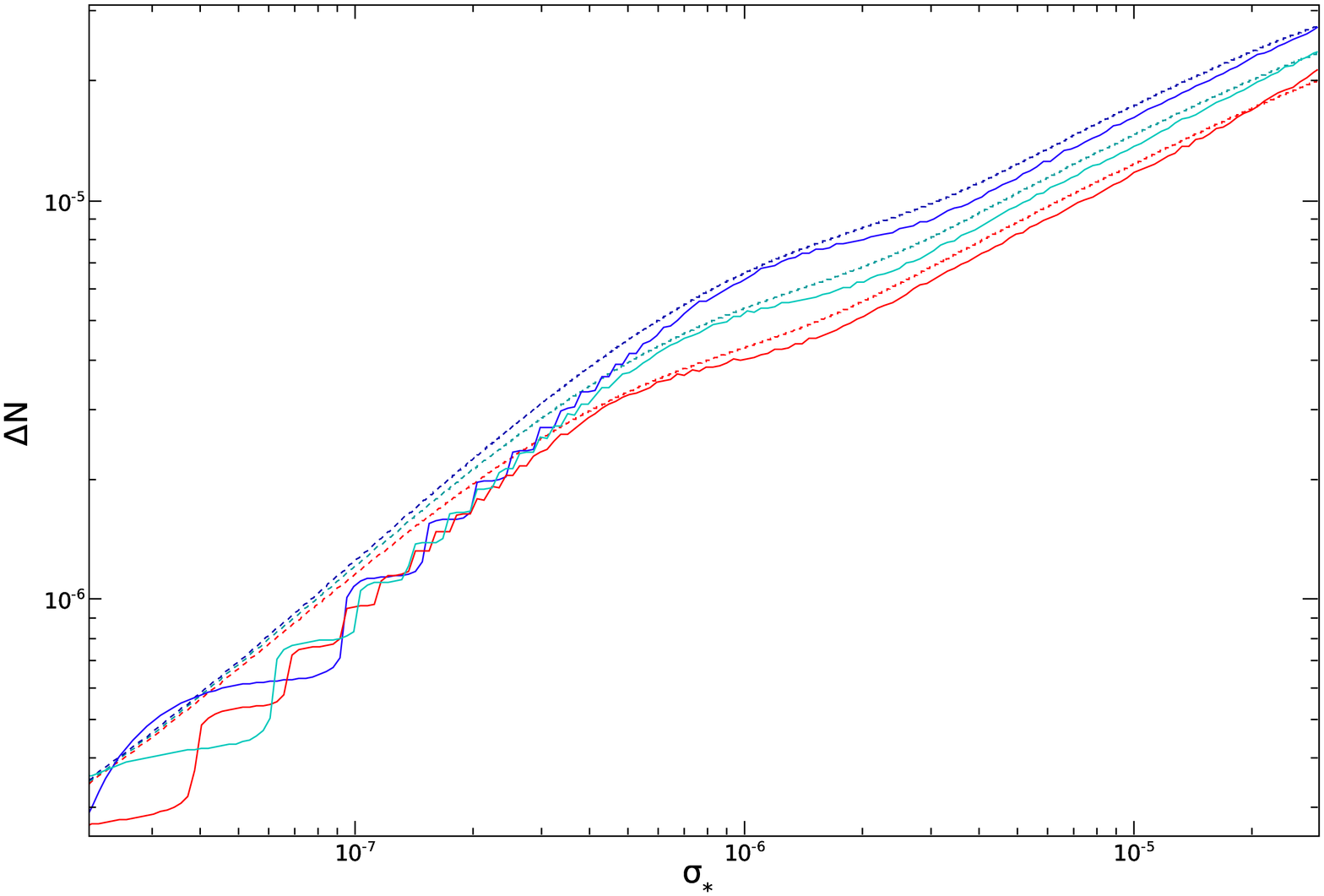}
 \caption{The curvature perturbation $\zeta$ for a
 choice of parameters $n=0$, $m=10^{-9}$, $H_{*}=10^{-7}$ and $\Gamma=10^{-28}$. The coupling values are $\lambda=10^{-7}$, $\lambda=2\times 10^{-7}$ and $\lambda=4\times 10^{-7}$
 for red, green and blue curves respectively. The solid lines show numerical results and
 dashed lines analytical estimates. }
 \label{fig:vertailu}
 \end{figure}
First, we can observe that the agreement between analytical and
numerical results is very good, even surprisingly so. The mismatch
seen for small field values occurs where the dynamics is dominated
by the quadratic potential $c\ll 1$ and the choice of initial
conditions for our analytical computation is no longer valid, as
discussed above. In this regime, we expect an error proportional to
$b\sim 1.2$ which seems to be well in line with the behaviour seen
in Fig. \ref{fig:vertailu}. The bump around $\sigma_{*}\sim 10^{-6}$
marks the regime where the quadratic and quartic terms are roughly
equal initially, $c\sim 1$, and the transition from quartic to
quadratic potential takes place soon after the onset of
oscillations. For larger field values, oscillations in the quartic
potential last for some while and the result can be understood using
the simple scaling law (\ref{scaling_law}). These features are
generic although discussed here using the specific choice of
parameters in Fig. \ref{fig:vertailu} as an example.

Unlike for the non-renormalizable potentials, we do not detect any
oscillatory behaviour of $\zeta$ in the regime $c\sim 1$ but instead
only the bump. Indeed, using (\ref{zeta_nb}) we have checked that
the first derivative of $\zeta$ with respect to $\sigma_{*}$ is a
positive definite function in the limit $r\ll 1$ and non-monotonous
features will show up only in third and higher order derivatives.
This differs qualitatively from the non-renormalizable case where
$\zeta$ oscillates as a function of $\sigma_{*}$ and consequently
even $\zeta'$ must change sign. Although we do not have an
analytical estimate valid for non-renormalizable potentials, this
agrees with general qualitative expectations as the transition into
the quadratic regime takes place earlier due to the more rapid
decrease of the field amplitude and is therefore more prone to leave
an imprint on the perturbation.

\begin{figure}[!h]
 \centering
 \includegraphics[scale=1.8]{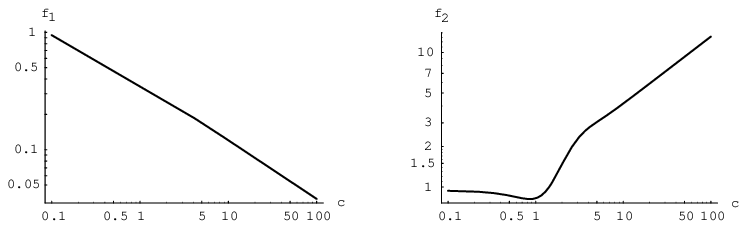}
 \label{fig:poikkeama}
 \caption{$f_1$ and $f_2$ plotted for different values of $c\sim 2\sqrt{\lambda}\sigma_*/m$.
 $f_1$ and $f_2$, defined in Eq.\ (\protect \ref{zeta_leading}),
 describe the deviation of $\zeta$ and $f_{\rm NL}$ from the purely
 quadratic results.}
 \end{figure}

Finally, using (\ref{zeta_exp}) we can express the amplitude of
Gaussian perturbations and the non-linearity parameter $f_{\rm
NL}=5/6 ({\rm ln}\,a)''/({\rm ln}\,a)'{}^2$ to leading order in $r$
and up to numerical factors in the form
  \beq
  \label{zeta_leading}
  \zeta\sim
  \sigma_{*}\left(\frac{m}{\Gamma}\right)^{1/2}\delta\sigma_{*}\times  f_1(c)\
  ,~~
  f_{\rm NL}\sim \frac{1}{\sigma_{*}^2}\left(\frac{\Gamma}{m}\right)^{1/2} \times f_2(c)\
  ,
  \eeq
which coincide with the well known quadratic results \cite{LUW}
apart from the functions $f_1=\sigma_{\rm osc}\sigma_{\rm
osc}'/\sigma_{*}$ and $f_2=(\sigma_*/\sigma_{\rm
osc})^2(1+{\sigma_{\rm osc}\sigma_{\rm osc}''}/ {\sigma_{\rm
osc}'{}^2})$. These depend only on the dimensionless parameter $c$
and their behaviour is depicted in Fig. \ref{fig:poikkeama}.
It can be seen that the inclusion of a quartic term in the potential
tends to decrease the perturbation amplitude and enhance the
non-Gaussian effects, except for the slight decrease\footnote{Note
that this differs from the result of \cite{kesn} based on a
perturbative analysis. For $n>0$, however, the sign change of
$f_{\rm NL}$ found in \cite{kesn} is consistent with the
oscillations of $\Delta N$ found in this paper.} of $f_{\rm
NL}$ observed around $c\sim 1$. However, deviations from quadratic
results are rather small as long as the quartic term is not vastly
dominant over the quadratic part during inflation. For example, it
can be seen from Fig. \ref{fig:poikkeama} that the quadratic results
get modified by at most an order of magnitude for $c\lesssim 10$,
where $c=10$ already corresponds to a clear dominance of the quartic
term. In this regime the model is therefore compatible with
observations under much of the same conditions as a quadratic model
with the same parameters $(\sigma_{*},H_{*},m,\Gamma)$, although one
also needs to take into account the different consistency
constraints coming from the masslessness and subdominance of
curvaton during inflation. If the interaction becomes more dominant,
$c\gtrsim 10$ or so, Fig. \ref{fig:poikkeama} implies that the main
constraints come from the observational upper bound $|f_{\rm
NL}|\lesssim 100$ for the bispectrum \cite{wmap}. This translates
into a lower bound for the curvaton subdominance $r$, which will be
more stringent than the quadratic result $r\gtrsim 0.01$ \cite{LUW}.
These limits from non-gaussianity need to be checked more carefully,
and calculated also for the cases $n > 0$ \cite{forth}.

\section{Conclusions}

We have studied the evolution of the amplitude of primordial
perturbation in a class of curvaton models where the potential
contains, in addition to the usual quadratic term, either a quartic
or some non-renormalizable term. For the calculation of the
perturbation amplitude, these turn out to be dynamically important
for a large part of the parameter space. We also focused on the
possibility that the curvaton energy density is subdominant
at the time of its decay. To some extent, this a natural consequence
of non-renormalizable curvaton potential since then the relative
curvaton energy density $r$ gets damped after inflation when the
field is oscillating in the non-quadratic part of its potential.
Depending on the decay time, this stage is followed by oscillations
in a quadratic potential whence $r$ starts to increase again, but
typically remains subdominant for realistic decay widths.

Although the amplification of curvaton energy is less efficient than
for a quadratic model, the interacting model may still generate the
observed amplitude of primordial perturbation since the initial
curvaton perturbation $\delta\sigma_{*}/\sigma_{*}$ can be much
larger than $10^{-5}$. This translates into constraints for the
model parameters: the decay width $\Gamma$, the inflationary scale
$H_*$, the bare curvaton mass $m$ and its initial energy density
$r_{*}$. In the present paper we have performed a systematic study
of the available parameter space for different types of
non-quadratic potentials. We have shown that the correct amplitude
of perturbations can indeed be achieved for large regions of the
parameter space even if the interaction term dominates over the
quadratic part and the curvaton would be subdominant by a factor of
$r \sim 10^{-3}$ or so at the time of its decay.

Moreover, we found that the final relative curvaton energy density
$r_{\rm decay}$ required to give rise to the observed amplitude depends on the
model parameters in a highly complicated way and, for
non-renormalizable potentials, shows oscillatory behaviour when the
non-quadratic part dominates initially. The same oscillatory features show up when considering the curvature perturbation $\zeta$ as a function of the curvaton
value $\sigma_{*}$ during inflation. Again, this happens for a large
portion of the parameter space. We have traced this non-monotonous
parametric dependence to the field dynamics taking place during the
transition from the non-quadratic to the quadratic part of the
potential. If the transition takes place soon after the beginning of
oscillations, both $\sigma$ and $\dot{\sigma}$ act as independent
dynamical variables and, consequently, variation of the initial
conditions can leave a non-trivial imprint on the expansion history.
If the non-quadratic term always remains subdominant such features are absent. Also for the renormalizable case $n=0$, the transition does not give rise to
oscillatory behaviour but nevertheless leaves a clearly distinguishable mark in
the results. This agrees with the analytical estimates we have
derived for this special case.

Some parts of the parameter space that give rise to an acceptable
perturbation amplitude are likely to be ruled out because they
produce large non-Gaussianities. This is expected as we consider the
subdominant limit $r\ll 1$ where the non-Gaussian effect are known
to be large already for a quadratic model. However, because of the
non-monotonic dependence $\zeta(\sigma_{*})$ observed for
non-renormalizable potentials, it is not evident what the final
non-Gaussianity turns out be in this case. Indeed, when considering
the expansion (\ref{def_zeta}) for the curvature perturbation, the
non-monotonic behaviour implies that the sign of $({\rm ln}\,a)''$
is varying, barring any accidental cancellations. In certain regions
of the parameter space we should therefore find a significant
suppression for the non-linearity parameter $f_{\rm NL}\sim ({\rm
ln}\,a)''/({\rm ln}\,a)'{}^{2}$ measuring the bispectrum amplitude.
It is clear though that this qualitative argumentation cannot be
pushed much further and a detailed survey of the parameter space is
required to address the question of non-Gaussianity. This will be
the topic of a forthcoming paper \cite{forth}.

\acknowledgments

This work was supported by the EU 6th Framework Marie Curie Research
and Training network "UniverseNet" (MRTN-CT-2006-035863) and partly
by the Academy of Finland grants 114419 (K.E.) and 130265 (S.N.).
O.T. is partly supported by the Magnus Ehrnrooth Foundation. T.T.
would like to thank the Helsinki Institute of Physics for the
hospitality during the visit, where this work was initiated. The
work of T.T. was supported by the Grant-in-Aid for Scientific
Research from the Ministry of Education, Science, Sports, and
Culture of Japan No.~19740145.


\begin{thebibliography}{999}

\bibitem{curvaton}
K.~Enqvist and M.~S.~Sloth,
\npb{626}{2002}{395}
\hepph{0109214};
%
D.~H.~Lyth and D.~Wands,
Phys.\ Lett.\ B {\bf 524}, 5 (2002)
\plb{524}{2002}{5}
\hepph{0110002};
%
T.~Moroi and T.~Takahashi,
Phys.\ Lett.\ B {\bf 522}, 215 (2001) [Erratum-ibid.\ B {\bf 539},
303 (2002)]
\hepph{0110096};
%
A.~D.~Linde and V.~F.~Mukhanov,
Phys.\ Rev.\  D {\bf 56} (1997) 535
\astroph{9610219};
%
S.~Mollerach,
Phys.\ Rev.\  D {\bf 42} (1990) 313.


\bibitem{mixed}
D.~Langlois and F.~Vernizzi,
Phys.\ Rev.\  D {\bf 70} (2004) 063522
\astroph{0403258};
%
G.~Lazarides, R.~R.~de Austri and R.~Trotta,
Phys.\ Rev.\  D {\bf 70} (2004) 123527
\hepph{0409335};
%
F.~Ferrer, S.~Rasanen and J.~Valiviita,
JCAP {\bf 0410} (2004) 010
\astroph{0407300};
%
T.~Moroi, T.~Takahashi and Y.~Toyoda,
Phys.\ Rev.\  D {\bf 72}, 023502 (2005)
\hepph{0501007};
%
T.~Moroi and T.~Takahashi,
Phys.\ Rev.\  D {\bf 72}, 023505 (2005)
\astroph{0505339};
%
K.~Ichikawa, T.~Suyama, T.~Takahashi and M.~Yamaguchi,
Phys.\ Rev.\  D {\bf 78}, 023513 (2008)
\arXivid{0802.4138}.

\bibitem{many_curvatons}
H.~Assadullahi, J.~Valiviita and D.~Wands,
Phys.\ Rev.\  D {\bf 76}, 103003 (2007)
\arXivid{0708.0223};
%
J.~Valiviita, H.~Assadullahi and D.~Wands,
\arXivid{0806.0623}.
%
%
\bibitem{LUW}
D.~H.~Lyth, C.~Ungarelli and D.~Wands,
Phys.\ Rev.\  D {\bf 67}, 023503 (2003)
\astroph{0208055}.

\bibitem{isocurvature}
T.~Moroi and T.~Takahashi,
Phys.\ Rev.\  D {\bf 66}, 063501 (2002)
\hepph{0206026};
%
D.~H.~Lyth and D.~Wands,
Phys.\ Rev.\  D {\bf 68} (2003) 103516
\astroph{0306500};
%
M.~Beltran,
Phys.\ Rev.\  D {\bf 78}, 023530 (2008)
\arXivid{0804.1097};
%
T.~Moroi and T.~Takahashi,
Phys.\ Lett.\  B {\bf 671}, 339 (2009)
\arXivid{0810.0189}.

\bibitem{alt_inflation}
See e.g.
L.~Kofman and S.~Mukohyama,
Phys.\ Rev.\  D {\bf 77} (2008) 043519
\arXivid{0709.1952}.


\bibitem{curvatondecres}
K.~Enqvist, S.~Nurmi and G.~I.~Rigopoulos,
JCAP {\bf 0810}, 013 (2008)
\arXivid{0807.0382}.

\bibitem{BasteroGil:2003tj}
M.~Bastero-Gil, V.~Di Clemente and S.~F.~King,
Phys.\ Rev.\  D {\bf 70} (2004) 023501
\hepph{0311237}.

\bibitem{kesn}
K.~Enqvist and S.~Nurmi,
JCAP {\bf 0510}, 013 (2005)
\astroph{0508573}.

\bibitem{kett}
K.~Enqvist and T.~Takahashi,
JCAP {\bf 0809}, 012 (2008)
\arXivid{0807.3069}.

\bibitem{curvaton_flat}
See e.g.
K.~Enqvist, A.~Jokinen, S.~Kasuya and A.~Mazumdar,
Phys.\ Rev.\ D {\bf 68}, 103507 (2003)
\hepph{0303165};
%
K.~Enqvist, S.~Kasuya and A.~Mazumdar,
Phys.\ Rev.\ Lett.\  {\bf 90}, 091302 (2003)
\hepph{0211147};
%
K.~Enqvist, S.~Kasuya and A.~Mazumdar,
Phys.\ Rev.\ Lett.\  {\bf 93}, 061301 (2004)
\hepph{0311224};
%
K.~Enqvist, A.~Mazumdar and A.~Perez-Lorenzana,
Phys.\ Rev.\ D {\bf 70}, 103508 (2004)
\hepth{0403044};
%
M.~Postma,
Phys.\ Rev.\ D {\bf 67}, 063518 (2003)
\hepph{0212005};
%
S.~Kasuya, M.~Kawasaki and F.~Takahashi,
Phys.\ Lett.\ B {\bf 578}, 259 (2004)
\hepph{0305134};
%
R.~Allahverdi,
Phys.\ Rev.\  D {\bf 70} (2004) 043507
\astroph{0403351};
%
M.~Ikegami and T.~Moroi,
Phys.\ Rev.\ D {\bf 70}, 083515 (2004)
\hepph{0404253};
%
R.~Allahverdi, K.~Enqvist, A.~Jokinen and A.~Mazumdar,
JCAP {\bf 0610} (2006) 007
\hepph{0603255}.

\bibitem{dynamics}
K.~Dimopoulos, G.~Lazarides, D.~Lyth and R.~Ruiz de Austri,
Phys.\ Rev.\  D {\bf 68} (2003) 123515
\hepph{0308015}.

\bibitem{Huang}
Q.~G.~Huang,
JCAP {\bf 0811} (2008) 005
\arXivid{0808.1793}.

\bibitem{curvaton_ng}
N.~Bartolo, S.~Matarrese and A.~Riotto,
Phys.\ Rev.\  D {\bf 69}, 043503 (2004)
\hepph{0309033};
%
C.~Gordon and K.~A.~Malik,
Phys.\ Rev.\  D {\bf 69}, 063508 (2004)
\astroph{0311102};
%
K.~A.~Malik and D.~H.~Lyth,
JCAP {\bf 0609}, 008 (2006)
\astroph{0604387};
%
M.~Sasaki, J.~Valiviita and D.~Wands,
Phys.\ Rev.\  D {\bf 74}, 103003 (2006)
\astroph{0607627};
%
J.~Valiviita, M.~Sasaki and D.~Wands,
\astroph{0610001}.

\bibitem{equilibrium}
A.~A.~Starobinsky and J.~Yokoyama,
Phys.\ Rev.\  D {\bf 50}, 6357 (1994)
\astroph{9407016}.

\bibitem{gradient_expansion}
D.~S.~Salopek and J.~R.~Bond,
Phys.\ Rev.\  D {\bf 42} (1990) 3936.


\bibitem{deltaN}
A.~A.~Starobinsky,
JETP Lett.\  {\bf 42} (1985) 152
[Pisma Zh.\ Eksp.\ Teor.\ Fiz.\  {\bf 42} (1985) 124];
%
M.~Sasaki and E.~D.~Stewart,
Prog.\ Theor.\ Phys.\  {\bf 95}, 71 (1996);
%
M.~Sasaki and T.~Tanaka,
Prog.\ Theor.\ Phys.\  {\bf 99}, 763 (1998).
%
%
\bibitem{recent_deltaN}
D.~Wands, K.~A.~Malik, D.~H.~Lyth and A.~R.~Liddle,
Phys.\ Rev.\  D {\bf 62} (2000) 043527
\astroph{0003278};
%
D.~H.~Lyth and D.~Wands,
Phys.\ Rev.\ D {\bf 68}, 103515 (2003)
\astroph{0306498};
%
D.~H.~Lyth, K.~A.~Malik and M.~Sasaki,
JCAP {\bf 0505}, 004 (2005)
\astroph{0411220};
%
D.~H.~Lyth and Y.~Rodriguez,
Phys.\ Rev.\ Lett.\  {\bf 95} (2005) 121302
\astroph{0504045}.

\bibitem{cobenorm}
C.~L.~Bennett {\it et al.},
Astrophys.\ J.\  {\bf 464} (1996) L1
\astroph{9601067}.

\bibitem{wmap}
E.~Komatsu {\it et al.}  [WMAP Collaboration],
Astrophys.\ J.\ Suppl.\  {\bf 180}, 330 (2009)
\arXivid{0803.0547}.


\bibitem{turner}
M.~S.~Turner,
Phys.\ Rev.\  D {\bf 28}, 1243 (1983).

\bibitem{GR}
See e.g.\
I.~S.~Gradshteyn and I.~M.~Ryzhik
``Table of Integrals, Series, and Products,''
Academic Press Inc.,Third Printing, 1969.

\bibitem{forth}
K.~Enqvist, S.~Nurmi, G.~I.~Rigopoulos, O.~Taanila and T.~Takahashi,
in preparation.

\end{thebibliography}
\end{document}